\documentclass[longauth]{aa}

\usepackage{graphicx}
\usepackage{txfonts}
\usepackage{orcidlink}
\usepackage{hyperref}
\usepackage[switch]{lineno}

\usepackage{multirow}
\usepackage{colortbl}

\usepackage{gensymb}
\usepackage{dirtytalk}

\newcommand{\HI}{\ion{H}{I}}
\newcommand{\NaI}{\ion{Na}{I}}
\newcommand{\NiII}{\ion{Ni}{II}}
\newcommand{\KI}{\ion{K}{I}}
\newcommand{\CaII}{\ion{Ca}{II}}
\newcommand{\OI}{\ion{O}{I}}
\newcommand{\FeII}{\ion{Fe}{II}}
\newcommand{\FeIII}{\ion{Fe}{III}}
\newcommand{\SII}{\ion{S}{II}}
\newcommand{\CoII}{\ion{Co}{II}}
\newcommand{\MgII}{\ion{Mg}{II}}
\newcommand{\SiII}{\ion{Si}{II}}
\newcommand{\ebv}{\textit{E}(\textit{B$-$V})}

\begin{document}

   \title{An analysis of the Type Ia SN~2024gy and a comparison of different host extinction estimation techniques}

   \author{Jacco H. Terwel \orcidlink{0000-0001-9834-3439} \inst{1}
          \and Kate Maguire \orcidlink{0000-0002-9770-3508} \inst{1}
          \and Cillian O'Donnell \orcidlink{0009-0007-5034-6420}\inst{1}
          \and Miika Pursiainen \orcidlink{0000-0003-4663-4300} \inst{2}
          \and Alba Casasbuenas \orcidlink{0000-0002-6677-3861} \inst{3, 4}
          \and Julie Thiim Gadeberg \inst{5}
          \and Ben Godson \orcidlink{0000-0003-3766-7266} \inst{2}
          \and Luke Harvey \orcidlink{0000-0003-3393-9383} \inst{1}
          \and Benjamin Nobre Hauptmann \orcidlink{0009-0009-0600-7581} \inst{6}
          \and Niilo Koivisto \orcidlink{0009-0007-7151-7313} \inst{7}
          \and Chang Liu \orcidlink{0000-0002-7866-4531} \inst{8, 9, 10}
          \and Shravya Shenoy \orcidlink{0009-0003-7025-089X} \inst{11}
          \and Samuel Grund S\o rensen \orcidlink{0009-0009-5887-4281} \inst{12}
          \and María Alejandra Díaz Teodori \orcidlink{0009-0002-1852-7671} \inst{7}
          \and Astrid Guldberg Theil \inst{12}
          \and Mikael Turkki \orcidlink{0009-0009-1581-1408} \inst{7, 13, 14}
          \and Alaa Alburai \orcidlink{0009-0007-2731-5562} \inst{15, 16}
          \and Joe Anderson \orcidlink{0000-0003-0227-3451} \inst{17}
          \and Thomas de Boer \orcidlink{0000-0001-5486-2747} \inst{18}
          \and Tomás Müller Bravo \orcidlink{0000-0003-3939-7167} \inst{1, 19}
          \and Umut Burgaz \orcidlink{0000-0003-0126-3999} \inst{1}
          \and Kenneth C. Chambers \orcidlink{0000-0001-6965-7789} \inst{18}
          \and Ting-Wan Chen \orcidlink{0000-0002-1066-6098} \inst{20}
          \and João Duarte \orcidlink{0000-0002-1823-3860} \inst{21}
          \and Lluis Galbany \orcidlink{0000-0002-1296-6887} \inst{15, 16}
          \and Mariusz Gromadzki \orcidlink{0000-0002-1650-1518} \inst{22}
          \and Cosimo Inserra \orcidlink{0000-0002-3968-4409} \inst{23}
          \and Joel Johansson \orcidlink{0000-0001-5975-290X} \inst{24}
          \and Young-Lo Kim \orcidlink{0000-0002-1031-0796} \inst{25}
          \and Thomas Lowe \orcidlink{0000-0002-9438-3617} \inst{18}
          \and Eugene Magnier \orcidlink{0000-0002-7965-2815} \inst{18}
          \and Rita P. Santos \orcidlink{0009-0002-2952-7431} \inst{17, 21}
          \and Jesper Sollerman \orcidlink{0000-0003-1546-6615} \inst{26}
          \and Richard Wainscoat \orcidlink{0000-0002-1341-0952} \inst{18}
          \and David R. Young \orcidlink{0000–0002–1229–2499} \inst{27}
          \and Tracy X. Chen \orcidlink{0000-0001-9152-6224} \inst{28}
          \and Matthew J. Graham \orcidlink{0000-0002-3168-0139} \inst{28}
          \and Mansi M. Kasliwal \orcidlink{0000-0002-5619-4938} \inst{29}
          \and Frank J. Masci \orcidlink{0000-0002-8532-9395} \inst{28}
          \and Josiah N. Purdum \orcidlink{0000-0003-1227-3738} \inst{30}
          \and Ines Belkhodja \orcidlink{0009-0006-4727-5361} \inst{29}
          }

   \institute{School of Physics, Trinity College Dublin, The University of Dublin, Dublin 2, Ireland,
   \email{terwelj@tcd.ie}
   \and Department of Physics, University of Warwick, Gibbet Hill Road, Coventry, CV4 7AL, UK
   \and Instituto de Astrofísica de Canarias, E-38200 La Laguna, Tenerife, Spain
   \and Departamento de Astrofísica, Universidad de La Laguna, E-38205 La Laguna, Tenerife, Spain
   \and Nordic Optical Telescope, Rambla José Ana Fernández Pérez 7, Breña Baja, La Palma 38711, Spain
   \and Institut de Física d'Altes Energies, Campus UAB, Facultat Ciencies Nord, 08193 Bellaterra, Barcelona
   \and Department of Physics and Astronomy, University of Turku, Vesilinnantie 5, Turku FI-20014, Finland
   \and Department of Physics and Astronomy, Northwestern University, 2145 Sheridan Rd, Evanston, IL 60208, USA
   \and Center for Interdisciplinary Exploration and Research in Astrophysics (CIERA), Northwestern University, 1800 Sherman Ave, Evanston, IL 60201, USA
   \and NSF-Simons AI Institute for the Sky (SkAI), 172 E. Chestnut St., Chicago, IL 60611, USA
   \and Centre for Astrophysics Research, University of Hertfordshire, Hatfield, AL10 9AB, 
   \and Department of Physics and Astronomy, Aarhus University, Ny Munkegade 120, 8000 Aarhus C, Denmark
   \and Finnish Centre for Astronomy with ESO (FINCA), University of Turku, FI-20014 Turku, Finland
   \and Aalto University Metsähovi Radio Observatory, Metsähovintie 114, 02540 Kylmälä, Finland
   \and Institute of Space Sciences (ICE, CSIC), Campus UAB, Carrer de Can Magrans, s/n, E-08193 Barcelona, Spain.
   \and Institut d’Estudis Espacials de Catalunya (IEEC), E-08034 Barcelona, Spain.
   \and European Southern Observatory, Alonso de Córdova 3107, Vitacura, Casilla 19001, Santiago, Chile
   \and Institute for Astronomy, University of Hawaii, 2680 Woodlawn Drive, Honolulu HI 96822
   \and Instituto de Ciencias Exactas y Naturales (ICEN), Universidad Arturo Prat, Chile
   \and Graduate Institute of Astronomy, National Central University, 300 Jhongda Road, 32001 Jhongli, Taiwan
   \and CENTRA, Departamento de F\'{\i}sica, Instituto Superior T\'ecnico -- IST, Universidade de Lisboa -- UL, Avenida Rovisco Pais 1, 1049-001 Lisboa, Portugal
   \and Astronomical Observatory, University of Warsaw, Al. Ujazdowskie 4, 00-478 Warszawa, Poland
   \and Cardiff Hub for Astrophysics Research and Technology, School of Physics \& Astronomy, Cardiff University, Queens Buildings, The Parade, Cardiff, CF24 3AA, UK
   \and Department of Physics, Oskar Klein Centre, Stockholm University, SE-106 91, Stockholm, Sweden
   \and Department of Astronomy \& Center for Galaxy Evolution Research, Yonsei University, Seoul 03722, Republic of Korea
   \and Oskar Klein Centre, Department of Astronomy, Stockholm University, Albanova University Center, SE-106 91, Stockholm, Sweden
   \and Astrophysics Research Centre, School of Mathematics and Physics, Queen’s University Belfast, Belfast BT7 1NN, UK
   \and IPAC, California Institute of Technology, 1200 E. California Blvd, Pasadena, CA, 91125, USA
   \and Division of Physics, Mathematics, and Astronomy, California Institute of Technology, Pasadena, CA 91125, USA
   \and Caltech Optical Observatories, California Institute of Technology, Pasadena, CA 91125, USA
   }

   \date{Received xxx; accepted yyy}

  \abstract
  {Type Ia supernovae (SNe Ia) are well-known standardisable candles, and are one of the main ways to measure the distance to their host galaxies. However, extinction due to interstellar dust causes objects to appear fainter and redder. Correcting for this requires estimating the amount of intervening material and how the extinction changes as a function of wavelength. We present and analyse optical and near-infrared data of the well-observed SN~2024gy and use these to compare different extinction estimation techniques, making use of photometric, spectroscopic, and polarimetric data. SN~2024gy is a normal SN Ia with high velocity (HV) components in \SiII~$\lambda6355$ (phase $<-10$ days) and a particularly strong HV feature in the \CaII\ near-infrared triplet (up to peak). Modelling SN~2024gy with \textsc{tardis} shows better matches with a double-detonation scenario compared to a delayed-detonation scenario due to a better match to the \CaII\ HV component. A measurement of the stable \ion{Ni}{}/\ion{Fe}{} ratio however favours a delayed-detonation scenario. Host extinction estimates range from \ebv$_{host}=0.12\pm0.02$ mag (narrow interstellar absorption lines) to \ebv$_{host}=0.24\pm0.06$ mag (Lira law) with a mean of \ebv$_{host}=0.22\pm0.04$ mag, assuming $R_V=3.1$. The spread between different methods highlights the challenge of accurately estimating the amount of extinction light suffers before being observed.}


   \maketitle

\section{Introduction}
\label{intro}
A white dwarf (WDs) in a binary system can become a Type Ia supernova (SN Ia) when the star is disrupted in a thermonuclear explosion. Normal events are well-known standardisable candles, as their peak absolute magnitude correlates well with their intrinsic colour and stretch \citep{Phillips_rel, colour_corr, Phillips_rel2}. This makes them excellent objects to measure distances and estimate the expansion rate of the universe with \citep[e.g.][]{Riess_accelerating_universe, Freedman_2001, Freedman_H0, Riess_SH0ES}.

The exact mechanisms that cause SN Ia explosions are still widely debated. It is generally accepted that the progenitor system is a binary consisting of the WD and another star, which may be degenerate \citep[Double degenerate scenario;][]{Iben_Double_degenerate, Webbink_Double_degenerate} or non-degenerate \citep[Single degenerate scenario;][]{Whelan_classical_Ia_mod, Nomoto_single_degenerate}. Both scenarios have several mechanisms that could ignite the WD and set off the explosion. In classical models, mass is transferred onto the WD until it gets close to the Chandrasekhar limit (M$_\text{Ch}\approx 1.4~\text{M}_\odot$, \citealt{Chandrasekhar_lim}) and carbon is ignited at the centre \citep{Whelan_classical_Ia_mod}. In delayed detonation models, the central carbon is burned during an initial deflagration phase before transitioning to a detonation \citep{Kholov_Del_det, Mazzali_common_mechanism}. 

In double-detonation models, an accreted layer of surface material ignites and explodes, sending shock waves into the WD. If this compresses the material near the core enough it can ignite and cause a second explosion \citep{Taam_ddet, Livne_ddet, Shen_ddet, Fink_ddet}. As the shock provides the temporary density and temperature spike that is needed to ignite the material, the WD in this scenario usually has a sub-Chandrasekhar mass. Some models have taken the double degenerate variant of this mechanism further, with the explosion of the primary igniting the surface material on the secondary WD and creating a second double detonation \citep{Pakmor_4dets, quadruple_det_Boos}. In other models, two WDs can collide or be involved in a violent merger which causes the explosion \citep{Rosswog_merger, Pakmor_merger, Pakmor_merger2}.

The interstellar medium (ISM) is not completely empty, and a fraction of the light from a SN (or any source) is absorbed or scattered by dust, which causes the SN to appear fainter. This effect is wavelength-dependent, with shorter wavelengths being more affected.  Correcting for extinction is vital to obtain accurate distance estimates to use in SN cosmology but this is not trivial and different parametrisations exist. Some of the most commonly used extinction functions include those presented in \citet{CCM89, O94, F99}, and \citet{FM_07}. Most of these use two parameters: $A_V$ is a measure of the extinction in the V band in mag, and $R_V$ is the total-to-selective extinction ratio such that \ebv$=A_V/R_V$ is the difference in extinction between the $B$ and $V$ bands. On average, $R_V=3.1$ for the Milky Way (MW), but $R_V$ depends on the grain size distribution, and MW regions with lower $R_V$ values exist \citep[e.g.][]{Udalski_low_MW_RV, Fitzpatrick_MW_extinction}. Studies have shown that the shape of the extinction curve, and therefore the average $R_V$ may be different in other galaxies \citep{14J_odd_Rv, RV_diffs_Amanullah, Brout_dust, GG_vary_col_lum_rel, Joel_extinction, Wiseman_Rv_step}.

Generally the properties of the material causing extinction cannot be measured directly, so one must rely on indirect tracers to estimate the total amount of extinction. For most SNe Ia, extinction is generally estimated by comparing multi-band photometry and spectroscopy to a sample of similar objects that are assumed to be mostly free of host extinction (e.g. by using SNe that are far from their hosts). However, if the SN is nearby and well observed, then high resolution spectra can be obtained, allowing for detection of narrow absorption features that can be attributed to the ISM or circumstellar material (CSM). 

One of the most well-used tracers of the ISM/CSM is the \NaI\ D doublet at 5890, 5896 \AA. \citet{NaID_rel_orig_paper} and \citet{KI_EBV_relation} found a relation between the \NaI\ D lines and MW \ebv\  while \citet{NaID_EBV_relation}  provided a more robust uncertainty estimation. Many studies extend this relation to extragalactic sources to estimate the amount of host extinction \citep[see e.g.][for some recent examples]{2020tlf, 23ixf_Smith, 2024ggi_Shrestha, 2022crv, 24pxl_Hoogendam, Iax_Magee, GRB_221009A}. Another common interstellar absorption doublet are the \CaII\ H\&K lines at 3934, 3969 \AA\ \citep{1979C_ISM_absorbers, SDSS_CaII_HK_quasars, neutral_gas_CaII_HK}. However, although the presence of these absorption features indicates the presence of gas and dust, no robust relation between the strength of these features and the amount of extinction has been derived \citep{Ca_II_bad_dust_estimator}.

If the spectral resolution is both high enough, there are other, fainter absorption lines that can be used to investigate the ISM and CSM along the line of sight. \citet{KI_EBV_relation} found a relation between \KI$_2$ 7699~\AA\ and the \ebv\ towards OB stars in the MW. \citet{Phillips_Na_overpredict_ebv} identified a correlation between the diffuse interstellar band (DIB) around 5780~\AA\ and extinction, and \citet{DIB_EBV_relation} found correlations for several MW DIBs and extinction for a large sample of extragalactic objects. These are due to absorption by unidentified molecules in the ISM \citep[see][for an overview]{DIB_overview, APO_DIB_cat} and can be used to study host galaxies of nearby SNe \citep[e.g.][]{Sollerman_DIBs}.

\citet{Li_24gy} present analysis of optical photometric and spectroscopic observations of SN~2024gy, a normal SN Ia in the nearby galaxy NGC~4216. They find that it has an anomalously strong high velocity (HV) \CaII\ near-infrared (NIR) feature in the earliest spectra, with velocities exceeding 25\,000~km~s$^{-1}$.  They also measure the \ion{Ni}{}/\ion{Fe}{} ratio in nebular spectra, finding it best matches with a delayed-detonation explosion scenario. \citet{Kwok_24gy} investigate SN~2024gy in the NIR and MIR using the \textit{James Webb Space Telescope} \textit{(JWST)} at late times, identifying a double-peaked emissivity profile of the Ni lines with an enhanced central emission, with the narrow core component having a slight redshift offset compared to the broader component. They connect this to a delayed-detonation model with a small off-centre ignition.

In this paper, we present ground-based optical photometric, spectroscopic, and polarimetric observations, and NIR spectroscopic data, of SN~2024gy. In Sect.~\ref{obs} the data is presented. In Sect.~\ref{lc}, \ref{spec}, and~\ref{pol} we analyse the photometry, spectroscopy, and polarimetry data, respectively. In Sect.~\ref{host_ext} we use different extinction estimation techniques to estimate \ebv$_{host}$. We discuss our findings in Sect.~\ref{discussion}, and conclude in Sect.~\ref{conclusion}.

\section{Observations and Data Reduction}
\label{obs}

\begin{figure}
    \centering
    \includegraphics[width=\columnwidth]{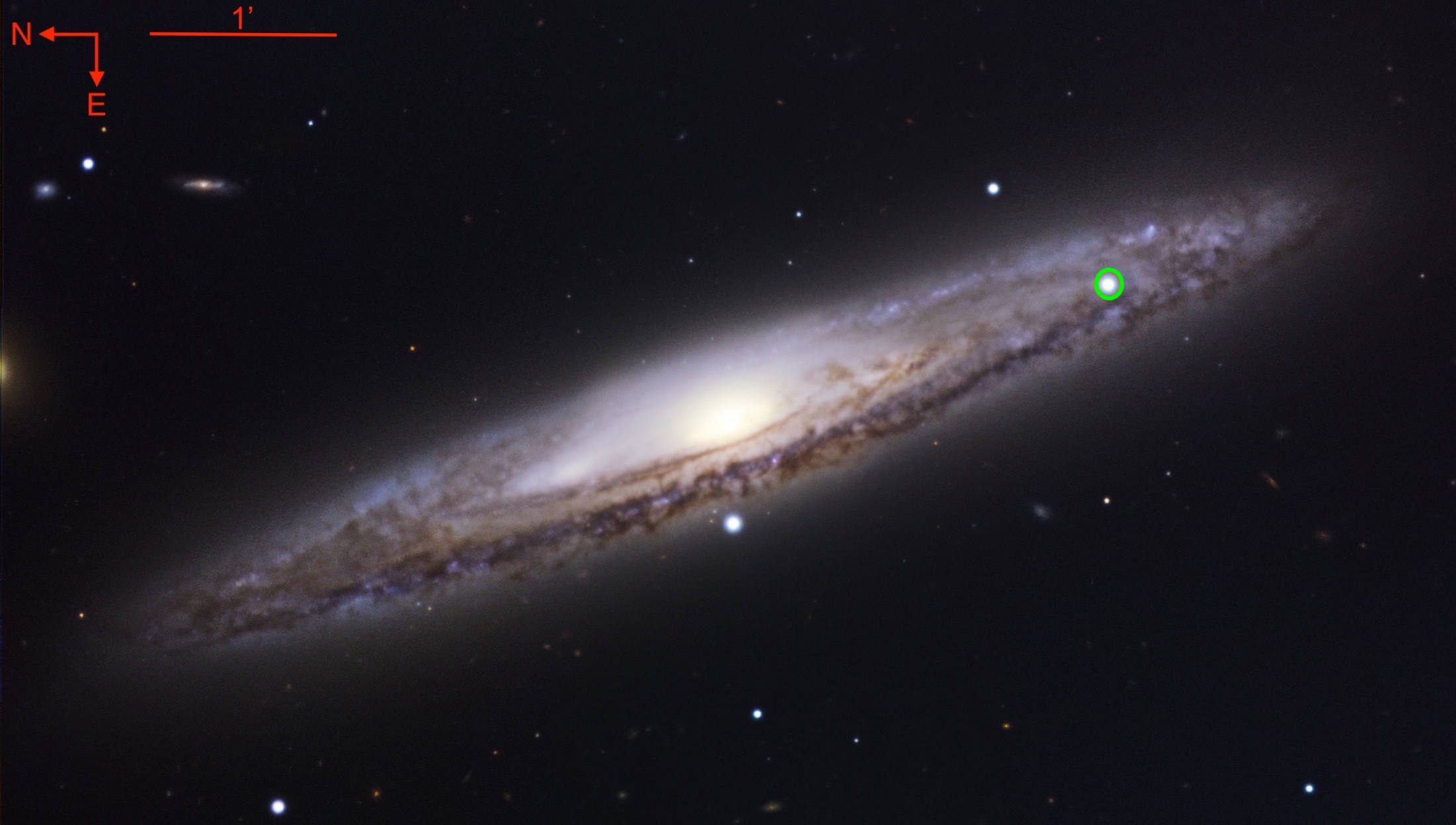}
    \caption{Colour image of NGC 4216 and SN~2024gy, taken with the NOT using the \textit{B}, \textit{V}, and \textit{R} filters. The SN is marked with a green circle.}
    \label{2024gy_NOT_colour}
\end{figure}

On 4 Jan 2024, the discovery of a new transient was reported to the Transient Name Server (TNS)\footnote{\url{https://www.wis-tns.org}} by Koichi Itagaki \citep{Itagaki_disc}. Located at R.A., Dec. = 12:15:51.289, +13:06:56.13, in NGC~4216, it was reported to have a Vega magnitude of $m=16.3$ in the clear filter. In the rest of this paper, we use AB magnitudes except where explicitly stated, and we assume the clear filter to be equivalent to the Landolt \textit{V} bandpass. The transient was classified the following day as a young Type Ia SN around two weeks before peak brightness at $z=0.001183$ \citep{2024gy_classification}, and was assigned the name SN~2024gy. The redshift was taken from a spectrum obtained by the Sloan Digital Sky Survey \citep[SDSS;][]{SDSS_tel, SDSS_DR13, SDSS_IV} at 5 arcsec from the SN location. We will use this redshift throughout the paper, as it falls in the range of redshifts we measure for host ISM absorption lines in our spectra (see Sect.~\ref{host_ext_spec}). \citet{Li_24gy} use the host redshift in their analysis, but do not take galaxy rotation at the SN location into account, which can have significant effects at low redshifts. Given its brightness and visibility, this SN provided an excellent opportunity for a long multi-wavelength follow-up campaign, which was performed with ground- and space-based telescopes and instruments. Figure~\ref{2024gy_NOT_colour} shows a colour image of SN~2024gy and NGC~4216, made using images taken with the Nordic Optical Telescope (NOT) in the \textit{B}, \textit{V}, and \textit{R} filters.

\subsection{Host galaxy}
The host galaxy of SN~2024gy, NGC~4216, is a massive edge-on barred spiral galaxy of type SAB(s)b \citep{deVaucouleurs_cat3}, with a prominent dust lane in its disk \citep{NGC4216_dustlane} at a redshift $z=0.000437\pm0.000013$, located on the outskirts of the Virgo cluster. The redshift difference between SN~2024gy and its host can be explained by a rotation velocity of about 200~km~s$^{-1}$ at the SN location \citep{NGC4216_rotation}. The galaxy does not show any signs of recent interaction or merger events, but deep imaging has shown an unusual amount of stellar streams and dwarf galaxies in the process of being disrupted and accreted \citep{NGC4216_dwarf_bombing}.

The distance to NGC~4216 has been measured many times in the literature. We adopt the Tully-Fisher \citep{Tully-Fisher_rel} distance found by \citet{NGC4216_dist} of a distance modulus $\mu=30.75\pm0.35$ mag, as it was the most recent measurement (older measurements are within $2\sigma$ from this value). According to the dust maps from \citet{SFD98_dust_maps}, the MW extinction in the direction of SN~2024gy is \ebv$_{MW}=$~$0.023$ mag after applying the correction found by \citet{SFD_offset} and \citet{SFD_recalib}. As suggested in these papers, we use the Fitzpatrick 1999 reddening law \citep{F99} when correcting for MW extinction throughout the paper.

The SDSS spectrum of NGC~4216 has visible H$\alpha$ and H$\beta$ lines. We can use these to get a first estimate of the host \ebv, though we note that it may not be representative as the spectrum was not taken at SN location in the galaxy. The intrinsic Balmer decrement (H$\alpha/$H$\beta$) at typical gas conditions in the galaxy can be calculated theoretically, and by comparing it to the observed value an estimate for the total extinction in the direction of the galaxy can be calculated. We take the EWs that are calculated by \textsc{astrodash} \citep{Astrodash}\footnote{\url{https://github.com/idies/SpecDash}}
and follow the procedure from \citet{Balmer_decrement_ebv} and find \ebv$_\text{tot}=0.27\pm0.16$.

\subsection{Photometry}
The location of SN~2024gy was observed by several sky surveys. We use these to create a rich multi-band optical photometry light curve covering the SN from before its explosion until after it faded below the detection thresholds of these surveys.

The All-Sky Automated Survey for Supernovae \citep[ASAS-SN;][]{ASASSN_paper1, ASASSN_paper2} $g$-band forced photometry was collected using the ASAS-SN Sky Patrol \citep{ASASSN_Skypatrol}. The Asteroid Terrestrial-impact Last Alert System \citep[ATLAS;][]{ATLAS,ATLAS_design_operation} \textit{c}- and \textit{o}-band forced photometry was obtained from their forced photometry web service \citep{ATLAS_FP_service}\footnote{\url{https://fallingstar-data.com/forcedphot/}}. The Gravitational-wave Optical Transient Observer \citep[GOTO;][]{GOTO_prototype, GOTO} images are processed with the \textsc{kadmilos} pipeline \citep{kadmilos}, and the \textit{L}-band forced photometry was performed using the GOTO Lightcurve service (Jarvis et al., in prep.). Forced \textit{iwy}-band photometry  from the Panoramic Survey Telescope \& Rapid Response System \citep[Pan-STARRS / PS;][]{Pan-STARRS1} was produced using a custom pipeline. The Zwicky Transient Facility \citep[ZTF;][]{ZTF_Surveys_Scheduler, ZTF_overview_and_1st_results, ZTF_Science_Objectives, ZTF_Instrumentation, ZTF_Observing_System} photometry comes from the \textsc{fpbot} package \citep{fpbot}~\footnote{\url{https://github.com/simeonreusch/fpbot}}, providing the ZTF~$g$, ZTF~$r$, and ZTF~$i$ bands. For ZTF, we removed data that had unphysical flux errors, poor PSF fits, were taken on cloudy nights, or had a failure in the image processing.

We also obtained 10 epochs of photometry in the $ugriz$ bands with the Infrared-Optical (IO:O) camera at the Liverpool telescope \citep[LT;][]{LT}. Template observations were taken with the same setup two years after the SN was discovered. $griz$ photometry was extracted using \textsc{autophot} \citep{autophot} with image subtraction using \textsc{pyzogy} \citep{zogy, pyzogy}. As NGC~4216 is very faint in the $u$ band at the location of SN~2024gy, we ran \textsc{autophot} without image subtraction for this band.

We rescale the flux to a common zero point ($zp=30$ mag), bin the data on a nightly basis, and correct for MW extinction. We then fit the light curves using the Spectral Adaptive Light Curve Template \cite[\textsc{salt2};][]{salt2}, as detailed in Section~\ref{lc}, to obtain the time of peak $B$-band brightness, $t_0 = 60329.7\pm0.1$, which we refer to as phase~0 throughout this work.

\subsection{Spectroscopy}
Table~\ref{speclog} lists the spectra presented in this paper with the phase of observation, the telescope and instrument used, and the covered wavelength range. Of the 28 spectra, 23 are in the optical, three spectra cover both the optical and NIR, and two spectra are in the NIR. The optical spectra are shown in shown in Fig.~\ref{all_optical_specs}. Most of these spectra have been collected through the Fritz broker \citep{skyportal2019, Skyportal}.  This dataset covers the SN at different stages of its evolution, containing 11 pre-peak spectra, three spectra within two weeks of peak, and 14 later spectra obtained up to 317 d post maximum light.

Ten spectra were obtained at Observatorio Roque de Los Muchachos in La Palma, Spain. Eight of these spectra were obtained at the NOT. Of these, six spectra were obtained with the Alhambra Faint Object Spectrograph and Camera (ALFOSC), one with the FIbre-fed Echelle Spectrograph \citep[FIES;][]{FIES_Telting}, and one with The NOT near-infrared Camera and spectrograph \citep[NOTCam;][]{NOTCam}\footnote{see also \url{https://www.not.iac.es/instruments/notcam/}}. One spectrum was obtained with the Intermediate Dispersion Spectrograph (IDS) at the Isaac Newton Telescope (INT), and one spectrum was obtained with the SPectrograph for the Rapid Acquisition of Transients \citep[SPRAT;][]{SPRAT} at the LT. The ALFOSC spectra were reduced using a custom data reduction pipeline based on \textsc{pypeit} \citep{pypeit:zenodo, pypeit:joss_arXiv, pypeit:joss_pub}~\footnote{\url{https://pypeit.readthedocs.io/en/latest/}}, which is available on GitHub~\footnote{\url{ https://github.com/steveschulze/NOT_DRP}}. The FIES spectrum was reduced using FIEStool, a Python based pipeline made available by the NOT. The NOTCam spectrum was reduced using standard IRAF routines. The IDS spectrum was reduced using standard IRAF/PYRAF and
IDL/Python routines, and the SPRAT spectrum was reduced using a custom Python-based pipeline.

Five spectra were public on WISeREP \citep{wiserep}~\footnote{\url{https://www.wiserep.org}}. One (the classification spectrum) was obtained with FLOYDS-N at the Faulkes Telescope North \citep{Faulkes}, operated by Las Cumbres Observatory. The other four spectra were obtained by amateur astronomer groups, who observed SN~2024gy with up to 1~m telescopes and used commercially available spectrographs. The continuum in the spectrum taken at a phase of $-1.4$~d is flatter than in the other spectra, suggesting a small issue with flux calibration. Otherwise the spectra look consistent with other spectra, with clear SN~Ia spectral features.

Four spectra were obtained at Mount Palomar. Three of these were obtained with the Spectral Energy Distribution Machine \citep[SEDm;][]{SEDM, SEDM_Kim} mounted on the Palomar 60-inch telescope \citep[P60;][]{P60} and reduced with \textsc{pysedm} \citep{pysedm}, and one spectrum was obtained with the Double Spectrograph (DBSP) mounted at the Palomar 200-inch telescope (P200) and reduced with \textsc{pyraf-dpbsp} \citep{pyraf_dbsp}. Two spectra were obtained with the Low Resolution Imaging Spectrometer \citep[LRIS;][]{LRIS} at the W.M. Keck observatory. Three spectra were obtained with the MMT, one with the MMT and Magellan Infrared Spectrograph \citep[MMIRS;][]{MMIRS}, and two  with Binospec. The latest two spectra had their host galaxy background removed using HostSub\_GP \citep{hostsub_gp}.

Two spectra were obtained with the EFOSC2 imaging spectrograph \citep{EFOSC2} on the ESO New Technology Telescope (NTT) in La Silla as part of the extended Public ESO Spectroscopic Survey of Transient Objects+ \citep[ePESSTO+;][]{PESSTO}. These spectra, observed with grisms 11 and 16, were reduced using the PESSTO pipeline~\footnote{\url{https://github.com/svalenti/pessto}} and are stitched together and presented as one in this paper. Finally, three epochs of spectra were obtained with X-Shooter \citep{X-Shooter} at the Very Large Telescope (VLT) and were reduced using the ESO X-Shooter pipeline. Each epoch consists of three spectra which are stitched together and presented as one in this paper, covering both the optical and NIR regime. We use \textsc{molecfit} \citep{Molecfit_II, Molefit_I} to correct the X-Shooter spectra for telluric absorption features.

\subsection{Polarimetry}
Besides photometry and spectroscopy, we also obtained $gri$-band linear polarimetry data at phases of $-8.5$, $-5.5$, $-1.5$, and $+50.5$ days relative to peak magnitude with ALFOSC at the NOT. The polarimetry was reduced using the bespoke pipeline presented in \citet{Pursiainen2025} following the reduction steps outlined in \citet{Pursiainen2023}. As the target is often elongated in the ALFOSC polarimetry mode images, we used large apertures of $r=2\times\mathrm{FWHM}$ to ensure all SN light is included. The data was taken using 16 angles at $22.5\degree$ intervals between $0\degree$ and $337.5\degree$. The reduction was performed in sets of four angles (e.g. $0.0\degree$, $22.5\degree$, $45.0\degree$ and $67.5\degree$), and these \say{cycles} were combined together for the final values presented in Sect.~\ref{pol}.

\section{Light curve properties}
\label{lc}
\begin{figure}
    \centering
    \includegraphics[width=\columnwidth]{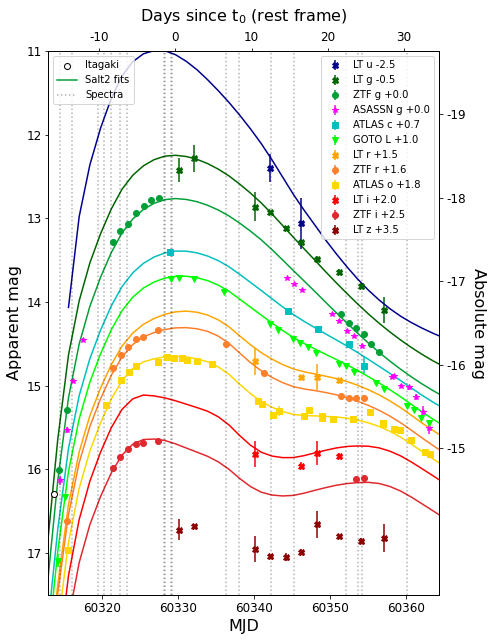}
    \caption{Light curves of the first 50 days of SN~2024gy, corrected for MW extinction but not for host galaxy extinction. The discovery is marked with a black circle. Four surveys observed SN~2024gy during this part of its evolution: ATLAS (squares, $c$ and $o$ bands are cyan and yellow, respectively), GOTO ($L$ band, lime inverted triangles), ZTF (circles, $g$, $r$, $i$ bands are green, orange, and red, respectively), and ASAS-SN ($g$ band, magenta stars). Additional photometry taken with the LT is also shown ($ugriz$ bands, coloured crosses). The points are plotted with a $1\sigma$ uncertainty, which in most cases is smaller than the marker. A vertical offset is added for readability, and the bands are ordered by their effective wavelength. The coloured lines show the \textsc{salt2} fits for each band except for ASAS-SN~$g$ as the observations do not cover enough of the light curve to be used in the \textsc{salt2} fits, and LT~$z$ as it is too red for \textsc{salt2}. The vertical dashed lines are the epochs at which a spectrum was taken.}
    \label{peak_lc}
\end{figure}

SNe Ia are typically characterised by the stretch and colour of their optical light curves around peak light. To find these values for SN~2024gy, we fit the ATLAS, GOTO, and ZTF observations of the first 50 days simultaneously using \textsc{SALT2} \citep{salt2} and show these in Fig.~\ref{peak_lc}. This fit gives a time of B~band maximum, $t_0 = 60329.676 \pm 0.003$ (phase~0) for SN~2024gy. We find a light curve width parameter of $x_1 = -0.216\pm0.004$  and a colour of $c=0.2105\pm0.0008$ mag. The stretch is fairly normal for a SN Ia. Compared to the full ZTF DR2 sample it is in the 40 percentile, meaning that SN~2024gy evolves slightly on the faster side \citep{DR2_lc}. The colour however is somewhat high for a SN Ia, with SN~2024gy being within the 15\% highest $c$ values in the ZTF DR2. Generally, such values indicate a non-negligible amount of extinction \citep[see e.g.][and references therein]{Brout_dust, Popovich_dust}.

At $\sim$25 days after peak, the light  curve reaches a secondary peak in the ZTF~$i$ band, as is typical for normal SNe Ia. In the ZTF~$r$ and ALTAS~$o$ bands this appears as a shoulder where the decline rate temporarily slows down. This feature is thought to be caused by the transition of iron-group elements from doubly to singly ionised as the ejecta cool down. The new ionisation state has comparatively more emissivity in the red and NIR, causing an increase in the flux in these regions \citep{Pinto2000, Kasen2006, Kasen2007, Dhawan2015}. \citet{Maxime_DR2} showed that the time of the ZTF~$i$ band secondary maximum correlates with the magnitude difference between peak and 15 days after the peak in the ZTF~$g$ band $\Delta m_{15}(g)$. Using the \textsc{salt2} fits, we calculate $\Delta m_{15}(g) = 0.82$~mag for SN~2024gy. A comparison with those SNe Ia shown in fig.~6 of \citet{Maxime_DR2} shows that the secondary peak in SN~2024gy is somewhat later than their fit predicts but falls within the bulk of the distribution, showing that SN~2024gy is broadly consistent with the ZTF DR2 sample in this regard.

\subsection{Pre-SN and post-SN binning}
\begin{figure*}
    \centering
    \includegraphics[width=0.97\textwidth]{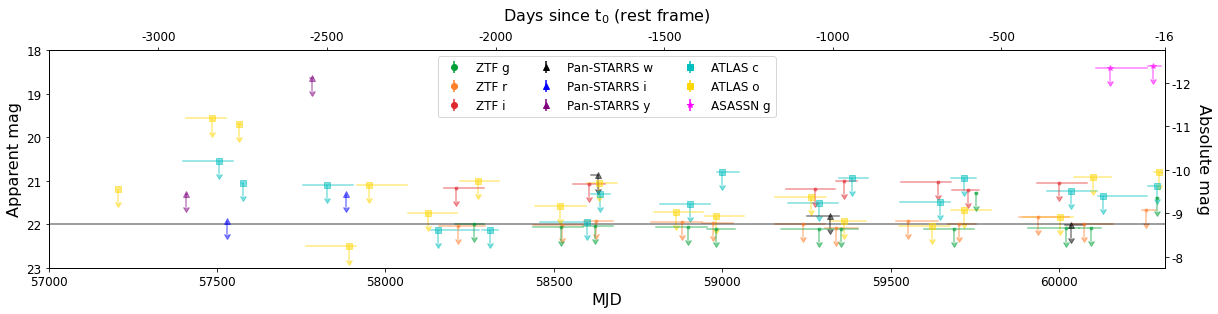}
    \includegraphics[width=0.97\textwidth]{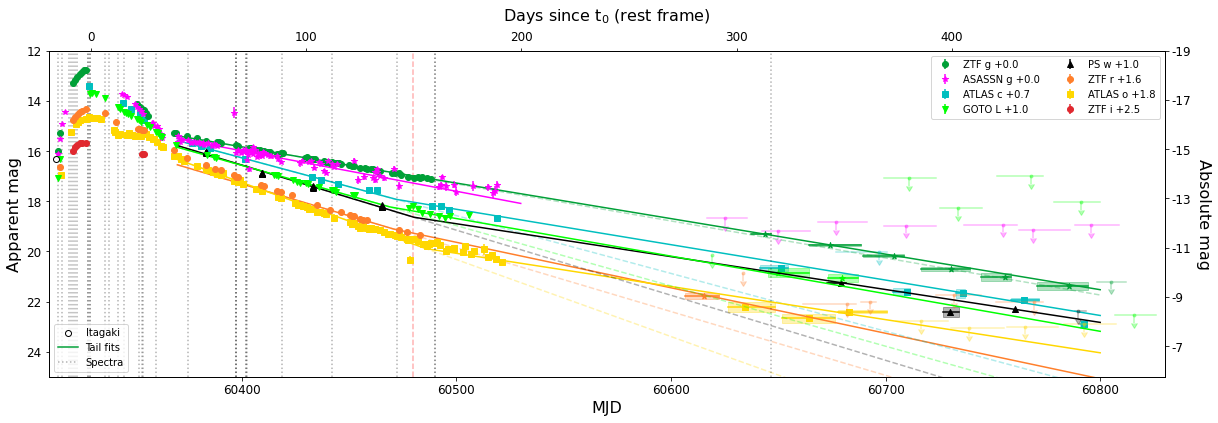}
    \caption{Full light curves of SN~2024gy, corrected for MW extinction but not for host galaxy extinction. The markers and colours are the same as for Fig.~\ref{peak_lc}, with the addition of Pan-STARRS data ($i$, $w$, and $y$ bands, blue, black, and purple triangles, respectively). As the LT observations have phases between 0 and 30 days after the peak they are not plotted here. The top panel shows the pre-discovery data in bins of up to 150 days. In cases where observations would only fill part of the bin (e.g.~the first 50 days) the bin edges have been moved to just encompass the binned data. The bottom panel shows the post-discovery observations. After the gap in the data, they are binned to recover the fading SN. These are shown with their width representing the bin size and a marker on the weighted mean MJD. Detections are shown with a $1\sigma$ uncertainty, and $5\sigma$ upper limits are shown with downward arrows for the non-detections. The lines show the tail fits for each band. The red vertical dashed line marks the switch from the first to the second tail fit. The other dashed lines extrapolate the first tail fit to late time, showing the impact of changing $t_{1/2}$.}
    \label{full_lc}
\end{figure*}

\begin{table}
    \centering
    \caption{Fitted tail parameters}
    \begin{tabular}{cccc}
        \hline
        \hline
         & \multicolumn{3}{c}{60370 $\leq t\leq 60480$} \\
        band & $a$ & $t_\text{ref}$ & $t_{1/2}$ (day) \\
        \hline
           ZTF \textit{g} & $6.434 \pm 0.004$ & 60370 & $51.0 \pm 0.1$ \\
           ZTF \textit{r} & $9.941 \pm 0.576$ & 60370 & $29.6 \pm 0.8$ \\
         ATLAS \textit{c} & $10.440 \pm 0.226$ & 60370 & $32.7 \pm 0.5$ \\
         ATLAS \textit{o} & $14.633 \pm 0.186$ & 60370 & $24.5 \pm 0.1$ \\
            PS \textit{w} & $11.579 \pm 0.404$ & 60370 & $29.0 \pm 0.9$ \\
          GOTO \textit{L} & $10.764 \pm 0.146$ & 60370 & $31.3 \pm 0.4$ \\
        ASAS-SN \textit{g}$^{(1)}$ & $5.802 \pm 0.117$ & 60370 & $46.6 \pm 1.5$ \\
        \hline
        \hline
         & \multicolumn{3}{c}{$t > 60480$}\\
        \hline
           ZTF \textit{g} & $1.472 \pm 0.023$ & 60480 &  $53.1 \pm 0.5$ \\
           ZTF \textit{r} & $0.806 \pm 0.001$ & 60480 &  $41.5 \pm 1.0$ \\
         ATLAS \textit{c} & $1.083 \pm 0.027$ & 60480 &  $53.3 \pm 1.2$ \\
         ATLAS \textit{o} & $0.594 \pm 0.019$ & 60480 &  $56.9 \pm 1.3$ \\
            PS \textit{w} & $0.836 \pm 0.073$ & 60480 &  $57.4 \pm 1.7$ \\
          GOTO \textit{L} & $1.035 \pm 0.024$ & 60480 &  $50.2 \pm 1.3$ \\
        \hline
    \end{tabular}
    \tablefoot{
    \tablefoottext{1}{As there are no detections in this filter after the observation gap, the second fit was not performed. The first fit is extrapolated to the last detection instead.}
    }
    \label{tail_fit_vals}
\end{table}

Figure~\ref{full_lc} shows the full optical light curves of SN~2024gy, with the top and bottom panels showing the pre- and post-discovery data, respectively. The big gap in observations between 200 and 280 days is because the SN was unobservable as it went behind the Sun.  As the SN fades, the signal falls below the single-image magnitude limit of the surveys. To follow the tail past this limit, we bin the observations taken after the gap in 25 day bins using the same method as \citet{JHT_DR2} to push the detection limit as deep as possible.

Starting at 40 days after the peak, we fit the observed flux up to 150 days after peak with a power law in each band. The tails are fitted in flux space with a decaying power law as 
\begin{equation}
    f(t)~=~10^5*a*2^{-(t-t_\text{ref})/t_{1/2}},
\end{equation}
with $a$ a scale parameter with ZP = 30 for conversion to magnitudes, $t$ the time in MJD, $t_\text{ref}$ a reference MJD at the start of the fitted region, and $t_{1/2}$ the half-life time of the power law in days. We do the same for the observations over 150 days after peak, effectively allowing for a kink in the SN tail at this phase. We allow for such a kink as previous studies of nearby SNe have shown the decaying tail to flatten out around this phase \citep[see e.g.][]{Zhang_11fe, Dimitriadis_11fe, JHT_DR2}. The fitted parameters are listed in Table~\ref{tail_fit_vals}.  These fits can be extrapolated to estimate the SN brightness at later phases, though these are lower limits past 600 days as another kink is expected there \citep{Dimitriadis_11fe}.

Large sky surveys have the advantage of observing the same part of the sky for years, which allows for the construction of a forced photometry light curve that extends years before the SN exploded and after it fades away. In the case of SN~2024gy, ZTF has nearly six years of observations before the SN exploded, while ATLAS and Pan-STARRS have observations dating back to July and December 2015, respectively. We searched the pre-SN observations for signs of precursor activity, pushing the detection limit beyond that of individual observations by binning the observations in 150 day bins. This is shown in the top panel of Fig.~\ref{full_lc}. We found no signs of precursor activity brighter than $m=22$ mag ($M=-8.75$, assuming no host extinction) in the approximate 9.5 years before the explosion.

\section{Spectral evolution}
\label{spec}
\begin{figure*}
    \centering
    \includegraphics[width=0.95\textwidth]{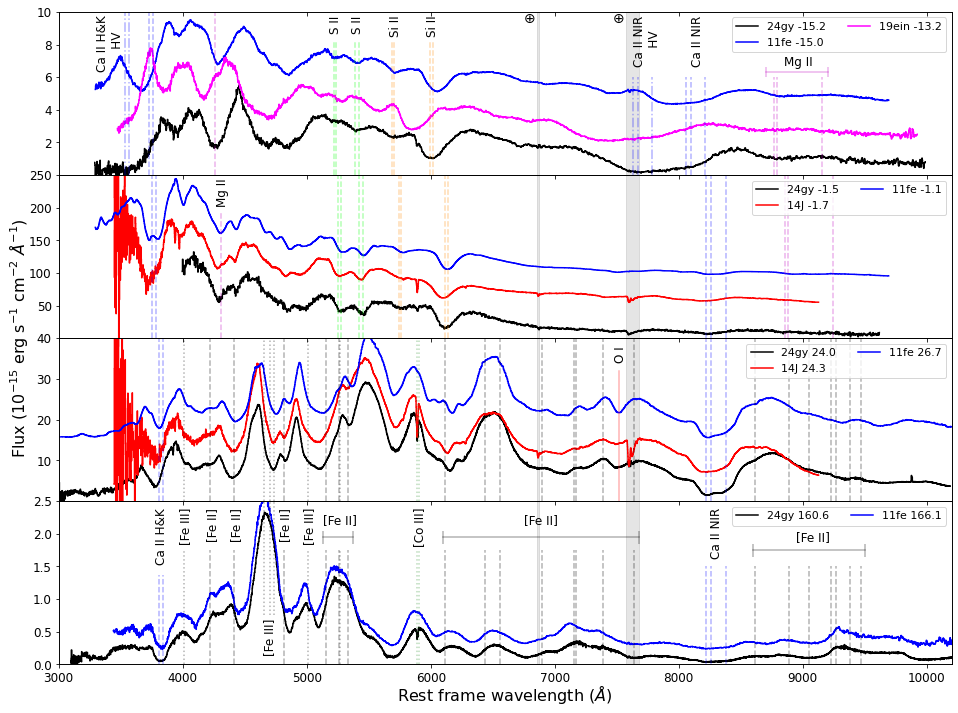}
    \caption{Optical spectra of SN~2024gy at four different phases of its evolution. Spectra of SN~2011fe, SN~2014J, and SN~2019ein at similar phases are shown with a vertical offset for comparison. All spectra are corrected for MW and host extinction, assuming $A_V=0.04$ mag and $R_V=3.1$ for the host extinction in SN~2024gy. Absorption and emission features are marked with vertical lines, with different line styles show different types of lines. Continuous lines are non-ionised, dashed lines are singly ionised, dotted lines are double ionised, and dot-dashed lines are high velocity components for \ion{Ca}{II}. Telluric bands are marked with grey bands. In the top two panels, various absorption lines are marked with the lines at the approximate location of the feature. In the bottom two panels we mark the emerging \ion{Fe}{} group element emission lines at rest wavelength which dominate the spectrum, but some absorption features from \CaII\ and \OI\ can still be seen at these phases.}
    \label{spec_showcase}
\end{figure*}

Figure~\ref{all_optical_specs} shows all optical spectra presented in this paper, while Fig.~\ref{spec_showcase} shows four spectra at different stages of the SN evolution compared with spectra of SN~2011fe \citep{Nugent_11fe, Li_11fe}, SN~2014J \citep{Goobar_14J}, and SN~2019ein \citep{2019ein_Kawabata}, and marks the dominant lines at each stage. SN~2011fe and SN~2014J were chosen as they are well-observed, nearby, normal SNe Ia, and we compare with SN~2019ein because of its prominent high velocity (HV) features. The spectra of SN~2011fe and SN~2014J from the Open Supernova Catalog\footnote{\url{https://github.com/astrocatalogs/supernovae}} \citep{Open_SN_cat}, and the SN~2019ein spectrum is from WISeREP. 

In the earliest spectrum of SN\,2024gy at $-$15.2 days with respect to peak brightness, we can see the \SiII\ lines at $\sim$12\,000--15\,000~km~s$^{-1}$ that define the SN Ia class. At the blue end the spectrum has a strong \ion{Mg}{II} line, and \CaII\ H\&K lines can be seen. On the red side, there is a broad absorption feature due to the \CaII\ NIR triplet, which is much broader than in SN~2011fe and requires a HV component to explain. We also mark the expected location of the corresponding \CaII\ H\&K HV lines, though we cannot fit the feature as it is at the edge of the spectrum. SN~2019ein displayed HV material at early stages, as can be seen by the main absorption features being more blueshifted in this object. While SN~2024gy generally matches better with SN~2011fe, the wide \CaII\ NIR feature looks more similar to that of SN~2019ein. Some \MgII\ lines can be seen at $\sim$9\,000 \AA\, and similar to the component at $\sim$4250 \AA, they match the velocity of the same features in SN~2011fe, while they have higher velocities in SN~2019ein.

Around peak the SN\,2024gy spectrum shows most of the same lines as in the first spectrum, but as the photosphere has receded further into the ejecta the absorption features have lower velocities of $\sim$10\,000~km~s$^{-1}$. The \CaII\ NIR feature has weakened in comparison to the rest of the spectrum, and the HV feature has disappeared. The spectrum of SN~2024gy is very similar to that of SN~2011fe and SN~2019ein. A few weeks after the peak, SN\,2024gy starts transitioning into the nebular phase. The \SiII\ feature is no longer visible (most likely due to line blending) and the spectrum starts to be dominated by emission from [\ion{Fe}{II}], [\ion{Fe}{III}], and [\ion{Co}{III}] lines. Above 6000 \AA, strong [\FeII] lines contribute, causing the SN light curve to show the secondary peak in the red and NIR regions. Prominent absorption features are also seen from \ion{O}{I}~$\lambda7772$ and in the \CaII\ NIR triplet and \CaII\ H\&K lines. Half a year after the SN exploded it is approaching the fully nebular phase and is dominated by emission from numerous \ion{Fe}{} group lines. Apart from the \CaII\ NIR triplet and \CaII\ H\&K lines, there are no absorption features present at this phase. Thereafter, the SN evolves more slowly as it cools and fades.

\subsection{Optical spectral line measurements}
In the following sections, we describe the measurements of the key observable features in the photospheric and nebular phases of the spectra of SN\,2024gy. We fit absorption features at photospheric phases, and fit forbidden emission lines at late time.

\subsubsection{Photospheric phase}
\begin{figure}
    \centering
    \includegraphics[width=0.98\columnwidth]{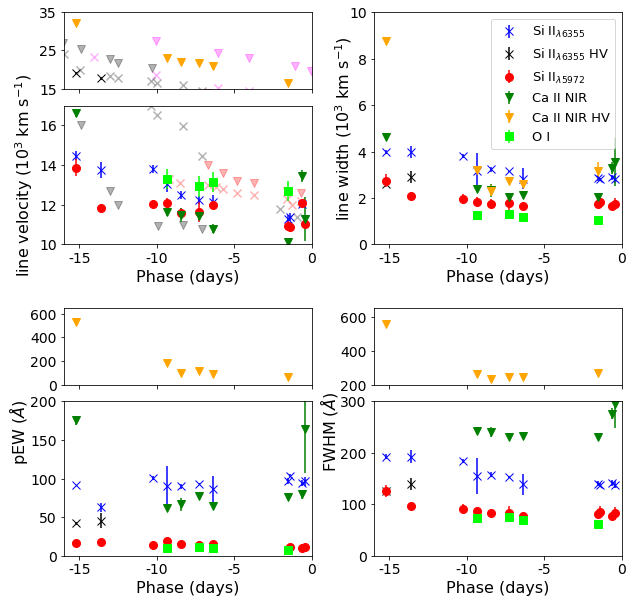}
    \caption{Measured line parameters in the pre-peak spectra. The HV components are plotted separately to increase readability. In the line velocity plot (top left) we also show in a lighter shade the \SiII~$\lambda6355$ (crosses) and \CaII\ (downward triangles) line velocities of SN~2011fe \citep[grey;][]{11fe_Parrent}, SN~2014J \citep[red;][]{14J_Marion}, and SN~2019ein \citep[magenta;][]{2019ein_pellegrino} for comparison. For SN~2019ein the \CaII\ H\&K lines were measured, for SN~2014J the \CaII\ NIR triplet was measured, and for SN~2011fe both features were measured.}
    \label{line_fits}
\end{figure}

The velocity evolution of absorption lines in the pre-peak spectra holds clues on the distribution of different elements in the ejecta. We fit the absorption features of the \SiII~$\lambda5972$ and \SiII~$\lambda6355$ doublets, \OI~$\lambda7222$ singlet, and the \CaII\ NIR triplet using a similar method as in \citet{Childress13} and \citet{Burgaz_DR2}. First, for numerical stability, we normalise the spectrum in the region we want to fit. Then we choose a small region on either side of the line and fit it with a straight line to estimate the pseudo-continuum, and subtract it from the spectrum. We then fit the continuum-subtracted spectrum with one Gaussian for each line in the complex and fix the relative strength, velocity offset and width of each Gaussian to be the same. This results in each absorption feature of a single element being fitted with a three-parameter model regardless of the number of Gaussians used. We allow for separate high-velocity components to be blended with normal velocity components \citep{Childress13, Maguire_PTF_spec_diversity, Silverman15, Burgaz_DR2, Harvey_DR2}, and use the Bayesian information criterion (BIC) fit statistic to select the best fit. To estimate the fit uncertainty we fit the model 10\,000 times while shifting the region used to estimate the background randomly up to $\pm4\,000$~km~s$^{-1}$. Finally, we add 200 km~s$^{-1}$ in quadrature to the line velocity uncertainties to account for the resolution of the our spectra and redshift uncertainty, as our adopted redshift was measured at a slightly different location in the host galaxy. The resulting line parameters of velocity, pseudo-equivalent width (pEW) and full width at half maximum (FWHM) are shown in Fig.~\ref{line_fits}.

We find two \SiII~$\lambda6355$ components in the two spectra at $-$15.2 and $-$13.6 days centred around 5950~\AA\ and 6050~\AA, respectively. Unfortunately, the following spectrum is too low resolution to clearly separate them and we fit it with a single component instead. At phases after $-$10.2 days, we no longer favour the addition of HV \SiII~$\lambda6355$ features based on the BIC statistic. While a HV component would also be expected for the \SiII~$\lambda5972$ line, the inclusion of such a component is not preferred at any epoch as the absorption feature is very weak. 

With over twice the width of any other measured line and a velocity of $>30\,000$~km~s$^{-1}$, the HV \CaII\ NIR at $-$15.2 days is the clear outlier. The spectrum at $-$13.6 days has a telluric residual overlapping with the \CaII\ NIR feature, preventing us from fitting it successfully. At $-$10.2 days, the HV feature has slowed down and become much narrower than it was in the first spectrum, though the HV feature remains strong. In the first spectrum, the broad \CaII\ NIR feature covers the region where we measure \OI~$\lambda7222$ later, but as the lower velocity regions are not visible yet at this epoch, we ignore possible line blending. Even if \OI\ is present at this phase, its contribution is likely minor. At later phases, when the HV \CaII\ has slowed down, we do detect some \OI\ absorption. But as the line is weak and in the same region as a major telluric line and potentially a \MgII\ line, fitting the line accurately remains difficult, and we only recover it in four spectra.

\citet{Li_24gy} also find a HV \CaII\ NIR component, though their maximum measured velocity is $\sim 5\,000$ km s$^{-1}$ lower. The difference in assumed SN redshift cannot account for this, but their first data point is a day after ours which could have given the component time to slow down. At phases after $-10$ days our measurements are consistent with those in \citet{Li_24gy}.

\subsubsection{Nebular phase}
The late-time spectra are dominated by emission lines from various elements, including many forbidden Fe-group lines. The emission feature between 7000~\AA\ and 7600~\AA\ in our latest spectrum (316.8 days after the peak) is a complex blend of four [\FeII] lines ($\lambda_{rest}=$ 7155, 7172, 7388, 7453~\AA) and two [\NiII] lines ($\lambda_{rest}=$ 7378, 7412~\AA). \citet{Maguire18} showed that this complex can be used to estimate the relative \ion{Ni}{} to \ion{Fe}{} abundance ratio using
\begin{equation}
    \frac{n_{Ni}}{n_{Fe}} = \frac{L_{7378}}{L_{7155}}\frac{d_{C_{Fe\ II}}}{d_{C_{Ni\ II}}}\frac{e^{-0.28/(kT)}}{4.9}.
    \label{eq:NiFe}
\end{equation}
where $n_{Ni}$ and $n_{Fe}$ are the number densities of \NiII\ and \FeII, respectively, $d_{C_{Fe\ II}}/d_{C_{Ni\ II}}$ is the ratio of departure coefficients of \FeII\ and \NiII\ from local thermodynamic equilibrium (LTE), which ranges between 1.2 and 2.4 at the phase we are interested in \citep{Jerkstrand15}. $k$ is the Boltzmann constant in units of eV K$^{-1}$, $T$ is the temperature in the range, 3\,000~K to 10\,000~K following \citet{Maguire18} and \citet{Flors_late_time_spec_progenitor}. $L_{7378}/L_{7155}$ is the measured flux ratio of the [\NiII]~$\lambda7378$ to [\FeII]~$\lambda7155$ line.

\begin{figure}
    \centering
    \includegraphics[width=\columnwidth]{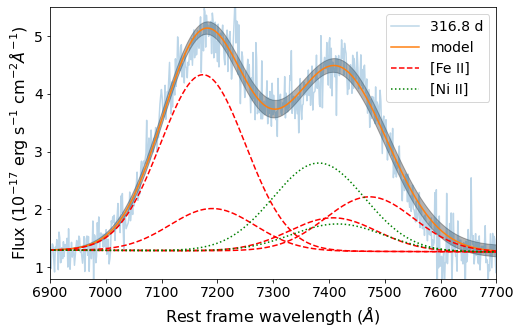}
    \caption{The fitted [\NiII]/[\FeII] complex and its separate components. The $1\sigma$ uncertainty on the fit is shown as a grey band around the fit.}
    \label{NiFe_fit}
\end{figure}

We fit the feature with a similar procedure as before, allowing each element to have different parameters. However, instead of fixing all amplitudes to be the same (as we did for the absorption lines in the pre-peak spectra) we use the relative strengths from \citet{Jerkstrand15}, giving the [\FeII] lines relative strengths of 1, 0.24, 0.1, and 0.31, respectively and the [\NiII] lines relative strengths of 1 and 0.31, respectively. The resulting fits are shown in Fig.~\ref{NiFe_fit}. The uncertainty on the fit is estimated by fitting the model 10\,000 times while shifting the background region randomly up to 1\,000~km~s$^{-1}$ for each fit. We calculate the pEWs of [\NiII]~$\lambda7378$ and [\FeII]~$\lambda7155$ from the fit and use this as an estimate for $L_{7378}/L_{7155}$. We put this estimate in Eq.~\ref{eq:NiFe} and draw 10\,000 values for the departure coefficient ratio and temperature from uniform distributions between the ranges listed above to get a mean and standard deviation for the abundance ratio. With this we find $n_{Ni}/n_{Fe}=0.11\pm0.05$, where the uncertainty is dominated by the 40\% uncertainty in the method \citep{Maguire18}. This is consistent with the $M_{Ni}/M_{Fe} = 0.080^{+0.053}_{-0.036}$ that was found for SN~2024gy by \citet{Li_24gy}.

\subsection{NIR spectra}
Along with the optical spectra, we also obtained four spectra covering the NIR up to $2.4\ \mu$m at phases of $-9.4$, $-$6.4, +15.6, and +67.7 days with respect to peak and one covering the \textit{J} band only at +24.5 days with respect to peak. These are shown in Fig.~\ref{NIR_specs} together with spectra of SN~2011fe and SN~2014J at similar phases for comparison. \citet{NIR_catalog} analysed an extensive catalogue of SN Ia NIR spectra, covering phases from 14 days before peak to 75 days after peak.  We use their list of absorption lines to identify features in SN~2024gy. The effect of extinction is much less at these wavelengths, and while we correct for MW extinction, the change is negligible.

The two pre-peak ($-$9.4 and $-$6.4 d) spectra are relatively featureless but some weak absorption features due to \MgII\ and \SiII\ are seen at velocities of $\sim$12\,000~km~s$^{-1}$. There is very good agreement between the spectra of SN~2024gy and those of SN~2011fe at these phases. A minor difference is the higher velocity of the 1.07 $\mu$m absorption feature seen for SN~2024gy compared to SN~2011fe. In some double-detonation models, this feature blends with \ion{He}{I}~$\lambda10830$ from unburnt material from the surface of the WD \citep{Dessart_He_det_mod, Boyle_He_dbldet, Collins_He_dbldet}, whose strength and velocity is highly dependent on viewing angle, which could explain the variations seen between the SNe, assuming this model is the correct one.

Post-peak spectra of SN\,2024gy were obtained at $+15.6$ d and $+24.5$ d ($J$ band only for the latter). At these phases more features are apparent in the spectra than in the first epochs. The main features come from hundreds of Fe-group lines, which start to dominate in certain regions through line-blanketing (grey bands in Fig.~\ref{NIR_specs}), increasing the opacity in these regions. This increases the effective photospheric radius at these wavelengths, raising the flux levels and mimicking emission features \citep{Wheeler_line_blanketing}. Gaps between the line-blanketing regions can also mimic absorption features. Overall, SN~2024gy looks very similar to SN~2014J at this stage, with only minor differences.

\begin{figure}
    \centering
    \includegraphics[width=0.98\columnwidth]{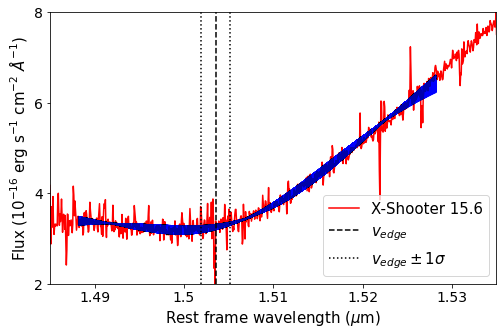}
    \caption{Fit of the $1.57\ \mu$m blue wing edge. The X-Shooter spectrum is shown in red, and the 10\,000 fits are shown in blue. The found location of $v_{edge}$ is marked with a dashed vertical line, and the $1\sigma$ uncertainty with dotted lines.}
    \label{NIR_vedge}
\end{figure}

The feature at $1.57\ \mu$m is a blend of \ion{Fe}{}, \ion{Co}{}, and \ion{Ni}{} lines, and while the centre of the feature is difficult to determine, the blue wing has a well-determined edge $\sim1.5\ \mu$m that is suggested to mark the outer edge of the synthesised $^{56}$\ion{Ni}{} \citep{Ashall_2019_NIR}. They showed that there is a significant difference in $v_{edge}$ between normal, transitional, and sub-luminous SNe. In \citet{Ashall_2019_phy} the measured $v_{edge}$ are compared to explosion models, and the sub-luminous SN~1999by is used as an example to show that $v_{edge}$ can be used to differentiate between $M_{Ch}$ and sub-$M_{Ch}$ explosion models. We follow the method used in \citet{Ashall_2019_NIR} to determine $v_{edge}$, assuming the rest wavelength to be $1.57\ \mu$m.  We first fit a region of $\sim0.05\ \mu$m around the blue wing edge with a straight line to determine the pseudo-continuum. After subtracting the pseudo-continuum we fit a Gaussian to the residual, and use its centroid to determine $v_{edge}$. The result is shown in Fig.~\ref{NIR_vedge}. As was done when fitting features in the optical, we fit this region 10\,000 times while shifting the background region randomly up to 1\,000 km s${-1}$. Using this method, we find $v_{edge}= -12400 \pm 300$ km s$^{-1}$. Unfortunately, we do not have models at this phase and wavelength range to compare against as is done in \citet{Ashall_2019_phy}.

The last NIR spectrum of SN\,2024gy was taken at 67.7 days after peak. At this stage the SN is in the transitional phase to becoming fully nebular, and the spectrum has evolved considerably compared to the previous ones. Line-blanketing from Fe-group lines continues to dominate large parts of the spectrum, causing broad features. Several absorption lines can be seen at velocities of 5\,000~km~s$^{-1}$, revealing the presence of \FeII, \MgII, and \CoII.

\subsection{TARDIS}
\label{tardis}
To gain insight into the potential explosion mechanism of SN2024gy, we perform spectral modelling with \textsc{TARDIS} \citep{tardis}. \textsc{TARDIS} is a one-dimensional radiative transfer code. It uses LTE approximations to calculate the ionization and excitation states, and is therefore not accurate when the ejecta becomes optically thin ($\sim$25 days from explosion). \textsc{TARDIS} takes an input abundance profile, time after explosion, luminosity, and photosphere velocity to generate a spectrum. We run TARDIS simulations for the double-detonation models presented in \citet{GRONOWDDMODELS}. These models are sub-M$_\text{Ch}$ mass CO WDs with a range of core masses ($0.8 < M_{core} < 1.1$) and \ion{He}{} shell masses ($0.02 < M_{He} < 0.1$). We also simulate the delayed-detonation model N100, which is an $\sim$M$_\text{Ch}$ CO WD \citep{N100MODEL}. We compare these model spectra to the X-Shooter spectra at $-8.5$~d and $-5.5$~d. 

Line-of-sight abundance profiles were used for the double-detonation models to preserve the asymmetry created by the \ion{He}{} detonation. Following the methodology of \citet{CALLANDOUBLEDET}, we took lines of sight from the North pole ($\theta = 0 \degree$), South pole ($\theta= 180\degree$), equator ($\theta = 90\degree$), as well as $\theta = 45 \degree$ and  $135\degree$. These models do not show great variation in the $\phi$ angle \citep{COLLINS2022}. The angle-averaged profile of N100 from \textsc{HESMA} \citep{HESMA} was used, as this model is more symmetric than the double-detonation models. For each model, we run a parameter grid in \textsc{TARDIS} and give each synthetic spectrum a quality score using the method described in O'Donnell et al. (in prep.).

\begin{figure}
    \centering
    \includegraphics[width=0.97\columnwidth]{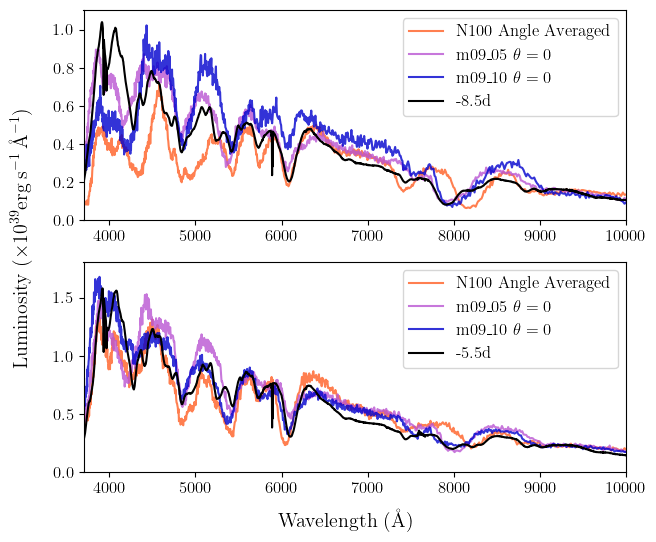}
    \caption{Best matching TARDIS simulations for the $-8.5$ d and $-5.5$ d epochs of SN 2024gy. At both epochs the N100 model is missing a high velocity Ca triplet feature.}
    \label{fig:tardis_match}
\end{figure}

\begin{figure}
    \centering
    \includegraphics[width=0.95\columnwidth]{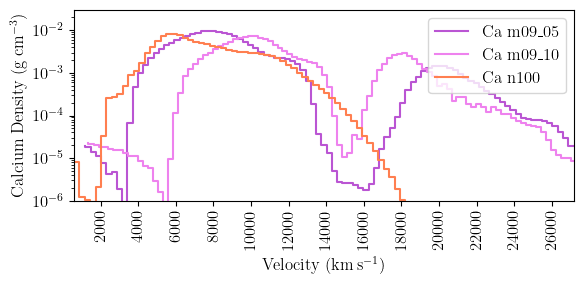}
    \caption{Profile in velocity space of the total density of calcium in m09\_05, m09\_10 and N100, showing the higher abundance of Ca at high velocities in the double-detonation models.}
    \label{fig:abund_profiles}
\end{figure}

Both epochs are reasonably well matched by the double-detonation and N100 models. The best match for the $-8.5$~d epoch comes from the \ion{He}{} detonation pole of m09\_05, ($0.9~M_\odot$ core and a $0.05~M_\odot$ He shell). For the $-5.5$~d this also matches well, but is second best to m09\_10 ($0.9~M_\odot$ core and a relatively large $0.1~M_\odot$ \ion{He}{} shell) along the same line-of-sight. The best matches for these models to each epoch are shown in Fig.~\ref{fig:tardis_match} and the corresponding parameters are given in Table~\ref{TARDIS_params}. A key difference between the double-detonation and N100 models is the absence of a HV feature of \CaII\ NIR triplet near $8000$~\AA\ in N100 and present in the double-detonation models due to the large abundance of \CaII\ produced at high velocities by the He-shell detonation (Fig.~\ref{fig:abund_profiles}). The double-detonation models are in better agreement with the observed feature. 

\begin{table}
    \centering
    \caption{Best matching \textsc{TARDIS} parameters for N100, and two double-detonation models, m09\_10 and m09\_05.}
    \resizebox{\columnwidth}{!}{
    \begin{tabular}{ccccc}
        \hline
        \hline
        Model & Phase & Luminosity (log$L_\odot$) & $t_{exp}$ (d) & $V_{inner}$ (km s$^{-1}$) \\
        \hline
        m09\_05 & $-$8.5 & 8.9 & 10.0 & 14000.0 \\
        m09\_05 & $-$5.5 & 9.1 & 13.0 & 13000.0 \\
        \hline
        m09\_10 & $-$8.5 & 8.9 & 10.0 & 10000.0 \\
        m09\_10 & $-$5.5 & 9.1 & 13.0 & 13000.0 \\
        \hline
        N100 & $-$8.5 & 8.8 & 12.0 & 12000.0 \\
        N100 & $-$5.5 & 9.1 & 12.0 & 12000.0 \\
        \hline
    \end{tabular}
    }
    \label{TARDIS_params}
\end{table}

\section{Polarimetry}
\label{pol}
Polarimetry is a powerful tool to observe asymmetries in SNe, and it has been used extensively to investigate asymmetric core-collapse SNe \citep[see][and references therein]{SN_specpol_overview}. SNe Ia on the other hand are generally assumed to be very spherical explosions overall, and little polarisation is expected in their continuum flux. However, as different explosion mechanisms predict viewing-angle dependent abundance profiles for various elements, SNe Ia are expected to show some model-dependent polarisation in different line features. Indeed, this has been found in a few cases such as SN~1996X \citep{1996X}, SN~1997dt \citep{1997dt_specpol}, SN~2001el \citep{2001el_specpol}, SN~2014J \citep{2014J_specpol}, SN~2018gv \citep{2018gv_specpol}, and SN~2021rhu \citep{2021rhu_specpol}, with the strongest signal generally coming from \SiII~$\lambda6355$ and the \CaII\ NIR triplet \citep{Cikota_Si_pol}.

\begin{figure}
    \centering
    \includegraphics[width=0.94\columnwidth]{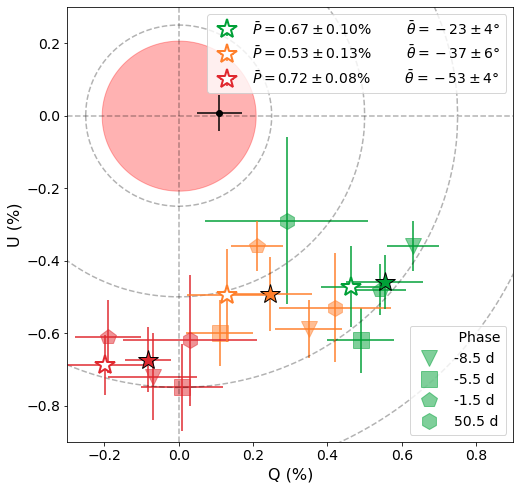}
    \caption{Stokes $Q$ and $U$ plane characterising the broadband polarisation of SN~2024gy at different phases. The $g$, $r$, and $i$ bands are shown in green, orange, and red, respectively. The filled stars mark the weighted mean of the total polarisation for each band, and the open stars mark the polarisation after correcting for MW polarisation. The other coloured markers show the values at each measured phase. The red shaded circle is $P_{max}$ that can be induced by MW dust alone according to the \citet{Serkowski_ISM_pol} upper limit, and the black point is our estimated value for the MW polarisation.}
    \label{impol}
\end{figure}

Figure~\ref{impol} shows the measured \textit{gri}-band imaging polarisation of SN~2024gy at each of the four phases where data was obtained ($-8.5$, $-5.5$, $-1.5$ and $+50.5$ days with respect to peak). The Stokes parameters describing polarisation are the degree of vertical to horizontal polarisation $Q$ and the degree of polarisation between the positive and negative diagonals $U$. From these, the total degree of polarisation $P=\sqrt{Q^2+U^2}$ and the position angle $\theta=\text{tan}^{-1}(U/Q)/2$ can be calculated, measured north over east relative to the north celestial pole. Note that $\theta$ is defined as the on-sky polarisation angle, which introduces the factor $1/2$ as one rotation in the $QU$ plane equals half a rotation on the sky.

We calculate the weighted mean polarisation in each band with and without the data at $50.5$ days as it has a much larger uncertainty. We find no significant change if we consider the pre-peak observations only, but the uncertainty reduces slightly. We find mean polarisations, $\bar{P_{g}}=0.72\pm0.08\%$, $\bar{P_{r}}=0.56\pm0.12\%$, and $\bar{P_{r}}=0.68\pm0.05\%$, and mean polarisation angles, $\bar{\theta_{g}}=-20\pm4\degree$, $\bar{\theta_{r}}=-32\pm5\degree$, and $\bar{\theta_{i}}=-48\pm4\degree$. There is no significant time evolution in any band, and the mean polarisation degree $\bar{P}$ is the same within the uncertainties for all bands. Only the polarisation angle $\theta$ changes between the bands.

The ISM can also induce a polarisation in the light that passes through. To isolate the polarisation that was induced in the host galaxy (SN + ISM) we need to subtract the MW polarisation in this direction. \citet{Serkowski_ISM_pol} showed that there is a well defined upper limit on the polarisation that the ISM can cause: $P_{max}(\%) = 9\times\ebv$. But to properly subtract the MW polarisation a value and direction is needed, not just an upper limit. To estimate this, we follow the method in \citet{Circinus_impol} and use the Stokes map from the Planck collaboration \citep{Planck15XXI, Planck18XII} to find the MW polarisation towards SN~2024gy at submm wavelengths. The degree of polarisation at submm and optical wavelengths are related through Eqs.~2 and 3 of \citet{Circinus_impol}, which gives us an estimate for the polarisation in the V~band. We can then convert this polarisation to the $g$, $r$, and $i$ bands using an empirical Serkowski law \citep{ISM_pol_review}. Using this method we find $P_\text{MW}=0.11 \pm 0.06\%$ and $\theta_\text{MW} = 2 \pm 13\degree$. The variation between the three bands is smaller than their uncertainties. Previous work has shown that this method gives comparable results to e.g. measuring the polarisation of nearby stars \citep{Circinus_impol, RV_impol_SNe}. As shown in Fig.~\ref{impol}, our $P_\text{MW}$ estimate is consistent with the \citet{Serkowski_ISM_pol} upper limit, and by subtracting $P_\text{MW}$ from the mean observed polarisation we find an estimate of the polarisation that the SN light had when it left the host galaxy.

SNe Ia are known to show high spherical symmetry, and our non-zero polarisation measurements are seemingly at odds with this. \citet{Cikota19} found an empirical relation between the maximum polarisation of \SiII~$\lambda6355$ (one of the most prominent polarised lines in SNe Ia together with the \CaII~NIR triplet) and its velocity at -5 days. Using their relation and the \SiII~$\lambda6355$ velocity of SN~2024gy at this phase ($\lesssim12000$ km s$^{-1}$) we find a maximum line polarisation of $P_{\SiII}\lesssim0.3\pm0.2\%$. On top of this, broadband filters are much wider than SN spectral features, decreasing the impact that the polarisation of these lines can have on the broadband polarisation. This suggests that the polarisation is not intrinsic to the SN, but induced by the ISM. \citet{Serkowski_ISM_pol} did a large multi-band investigation of MW ISM induced polarisation of starlight. They found a wavelength dependence in the polarisation angle for several stars that could be explained by the light passing through several clouds with different properties and dust grain orientations. This could also explain the wavelength dependence of the polarisation angle in SN~2024gy, as Sect.~\ref{ISM_features} we find evidence of two resolved host galaxy ISM clouds in the line-of-sight.

\section{Host extinction estimates}
\label{host_ext}
While the observed colour of a SN Ia is due to both intrinsic variations between SNe and external extinction, for SNe Ia with measured SALT2 values of $c>0.2$ mag, extinction is likely the dominant factor \citep[see e.g.][and references therein]{Brout_dust, Popovich_dust}. For SN\,2024gy, a value of $c=0.2105\pm0.0008$ mag was obtained in Sect.~\ref{lc}. This is consistent with the location of SN~2024gy likely being in a spiral arm in NGC~4216, where a non-negligible amount of host extinction is likely. Given our rich dataset of SN~2024gy, there are several methods through which we can estimate the \ebv$_{host}$, from both photometry (Lira law, light curve fitting, Sect,~\ref{host_ext_phot}) and spectroscopy (Sect.~\ref{host_ext_spec}). We also estimate a lower limit using the polarimetry measurements (Sect.~\ref{host_ext_pol}). The extinction estimates for all methods in this paper are listed in Table~\ref{reddening_estimates}.

\begin{table}
    \centering
    \caption{Host galaxy dust parameter estimates for SN~2024gy.}
    \resizebox{\columnwidth}{!}{
    \begin{tabular}{ccccc}
        \hline
        \hline
        Method & $A_V$ (mag) & $R_V$ & E$($B$-$V$)_\text{host}$ (mag)\\
        \hline
        Lira law & $0.7\pm0.2$& 3.1 & $0.24\pm0.06$\\
        \multirow{2}{*}{BayeSN ZTF} & $0.6 \pm 0.1$& 2.659 & $0.22 \pm 0.04$\\
         & $0.7 \pm 0.1$& 3.1 & $0.21 \pm 0.04$\\
        \multirow{2}{*}{BayeSN ZTF+ATLAS+GOTO} & $0.53 \pm 0.09$& 2.659 & $0.20 \pm 0.04$\\
         & $0.6 \pm 0.1$& 3.1 & $0.19 \pm 0.04$\\
        \multirow{2}{*}{Spectral matching} & $0.55 \pm 0.06$& 2.659 & $0.21 \pm 0.02$\\
         & $0.75 \pm 0.06$& 3.1 & $0.24 \pm 0.02$\\
        Mean of methods& $0.7 \pm 0.1$& 3.1& $0.22 \pm 0.04$ \\
        \hline
        \NaI\ D$^{(1,2)}$ & $3.8 \pm 0.5$ & 3.1 & $1.24 \pm 0.15$ \\
        \KI$_2$$^{(1)}$ & $0.80 \pm 0.27$& 3.1 & $0.26 \pm 0.09$ \\
        DIB$_{\lambda5780}$$^{(1)}$ & $0.36\pm0.03$ & 3.1 & $0.12\pm0.01$ \\
        DIB$_{\lambda5797}$$^{(1)}$ & $0.34\pm0.12$ & 3.1 & $0.11\pm0.04$ \\
        DIB$_{\lambda6613}$$^{(1)}$ & $0.7\pm0.4$ & 3.1 & $0.21\pm0.12$ \\
        Mean ISM lines$^{(3)}$ & $0.38\pm0.06$ & 3.1 & $0.12\pm0.02$ \\
        \hline
        Polarimetry & $\geq0.24\pm0.03$ & 3.1 & $\geq0.08 \pm0.01$ \\
        \hline
    \end{tabular}
    }
    \tablefoot{Each method assumes a value for $R_V$ and either estimates $A_V$ or \ebv, the other value is calculated by using $A_V = R_V\times\ebv$.\\ 
    \tablefoottext{1}{Weighted average of multiple measurements on the FIES and/or X-Shooter spectra (See Section~\ref{host_ext_spec}). If multiple clouds were found to contribute to the host extinction, they have been combined.\\}
    \tablefoottext{2}{The estimator using the combined \NaI\ D$_1$ + \NaI\ D$_2$ EWs is used here. Using the separate estimators for each lines gives results that agree within $1\sigma$ but have a larger uncertainty.\\}
    \tablefoottext{3}{\NaI\ D was excluded from this mean.\\}
    }
    \label{reddening_estimates}
\end{table}

\subsection{Photometric methods}
\label{host_ext_phot}
\paragraph{Lira Law.}
While SNe Ia can have intrinsic colour differences around peak, \citet{Phillips_rel2} showed that the (B-V) colour evolution of SNe Ia follows a linear decline (becoming bluer) between $\sim30-90$ days past B-band (or optical) maximum. This decline rate (or slope) is very similar for all SNe Ia and is called the `Lira Law' \citep{Lira_thesis, Phillips_rel2}. Any observed redder colour during this phase is most likely due to dust extinction. Therefore, if a SN is observed at these phases, the Lira law can be used to estimate the amount of extinction for a SN by comparing the colour curve to a extinction-free sample.

\begin{figure}
    \centering
    \includegraphics[width=0.95\columnwidth]{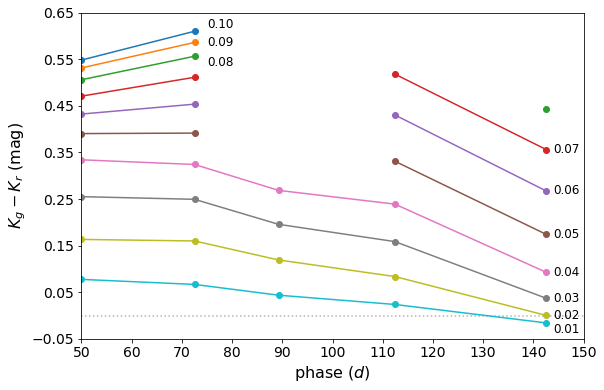}
    \caption{K-corrections at late phases for SN~2024gy at different redshifts. The y-axis is plotted as the difference in $g$ and $r$ K-correction, such that the corrected and observed colours are related as $(g-r)_{cor}=(g-r)_{obs} - (K_g-K_r)$. At 50 days we plot the K-corrections that are calculated with the \textsc{salt2} model. The other points use late-time SN~2024gy spectra shifted to different redshifts. Some of these spectra do not extend far enough into the blue to calculate $K_g$ at higher $z$, causing gaps in the plotted lines.}
    \label{Kcor_z}
\end{figure}

We compare the ZTF $g-r$ colour curve of SN\,2024gy to that of a sample of MW-extinction corrected normal SNe Ia from the ZTF SN Ia Data Release 2 \citep[ZTF DR2;][Smith et al.~in prep., which covers all ZTF observed SNe Ia that were first detected before 2021]{DR2_Overview}. The ZTF SNe for comparison where chosen to have $d_{DLR}\geq2$ \citep[directional light radius;][SN-galaxy separation in units of galaxy radius in the direction of the SN]{Sullivan_ddlr, Smith_ddlr, Gupta_ddlr} so that we have very low levels of expected extinction. To ensure we only use normal SNe with good \textsc{salt2} fits and low amounts of extinction, we only use objects with $x_1\geq-2.2$, $c\leq0.2$ \citep{DR2_lc}, \ebv$_{MW}\leq0.1$, and are subtyped as normal. To avoid selection biases, we only use the volume-limited DR2 sample ($z\leq0.06$). While this reduces the number of objects we can use, the removed objects are generally too faint to have detections at phases where the Lira law can be used. These cuts result in 40 normal SNe Ia with observations between 30 and 90 days after the peak that we assume to be free of host galaxy extinction.

We apply K-corrections to all objects using a standard \textsc{salt2} template. This template only goes up to 50 days after peak, so for later phases we use the K-correction at 50 days. To estimate the validity of this assumption, we calculate the K-corrections at various distances up to $z=0.1$ using our SN~2024gy spectra at phases between 30 and 150 days and compare this to the K-corrections given by the \textsc{salt2} model at 50 days. This is shown in Fig.~\ref{Kcor_z}. While the K-corrections do change over time at these phases, the shift is similar across all tested redshifts, with differences being $\lesssim0.05$ mag during the phases where the Lira law applies. The difference increases for larger $z$ but as these SNe were likely too faint to be detected at these phases, this part of the colour curve is mainly populated by nearby events.

The effect of extinction is an offset along the colour axis in the colour curve. Classically, the Lira law method works by fitting a straight line $c=a(t_p-60)+b$ to the colour $c$ at phases $30\leq t_p \leq90$ days for both the SN and the comparison sample. The difference in $b$ is directly related to the extinction in that colour. Since we use $g-r$ we get $E(g-r)$, which is the difference in extinction between the $g$ and $r$ bands. By assuming a dust law and $R_V$ we can then convert this to a measure of \ebv\ and $A_V$.

\begin{figure}
    \centering
    \includegraphics[width=0.95\columnwidth]{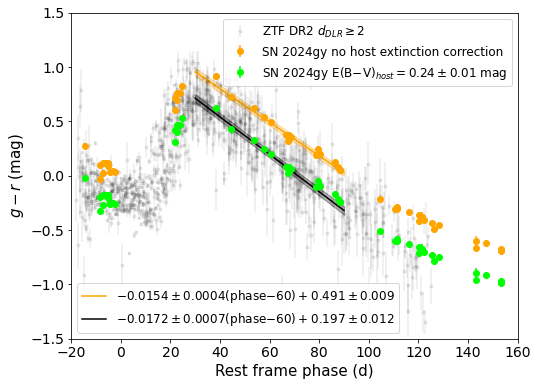}
    \caption{ZTF $g-r$ colour curve of SN~2024gy compared to the host dust free ZTF DR2 sample. All objects are corrected for MW extinction and have K-corrections applied. The orange line is the Lira law fit to SN~2024gy (orange points), and the black line is the Lira law fit to the comparison sample. Correcting SN~2024gy for the found host extinction results in the green points.}
    \label{gr_colour}
\end{figure}

Figure~\ref{gr_colour} shows the $g-r$ colour evolution of SN~2024gy and the comparison sample. The spread in the comparison sample at early phases are due to genuine colour differences between objects. At later phases, where the Lira law applies, the observed spread is due to larger uncertainties in the colour estimation caused by detections being closer to the ZTF detection limit. Fitting a line between $30-90$ days however still gives a tight relation to which the colour evolution of individual objects agree. We also fit a line to SN~2024gy at these phases and use this to find $E(g-r)=0.30\pm0.03$ mag. Assuming a \citet{F99} extinction law with $R_V=3.1$, we find $A_V=0.74\pm0.02$ mag and  \ebv$_\text{host}=0.24\pm0.01$ mag. Unfortunately, $A_V$ and $R_V$ are degenerate since we only have one colour to compare against; the ZTF $i$ band is too sparsely observed.

\citet{MW_Rv_map} present a detailed map of MW $R_V$ in different directions. While they find a mean value close to the often used $R_V=3.1$, it can vary between $2.4 \lesssim R_V \lesssim 4.2$. While higher galactic latitudes tend to have lower amounts of extinction, the $R_V$ can vary drastically along small angular separation. \ebv$_{MW}$ is low for SN~2024gy meaning this error source is subdominant, but this is not necessarily true for the comparison sample. We required \ebv$_{MW}\leq0.1$ mag, but the possible variation in $R_V$ can still allow the $g-r$ colour to vary up to 0.03 mag. Combining the uncertainty sources (K-correction, fit, and $R_V$) gives a final Lira law host extinction estimate of \ebv$_\text{host}=0.24\pm0.06$ mag.

\paragraph{BayeSN.}
Another way to estimate the host extinction is by fitting the light curve with BayeSN \citep{BayeSN_M20, BayeSN_Grayling}. We use the model trained by \citet{BayeSN_W22}, which by default assumes an $R_V=2.659$ but we also fit by assuming $R_V=3.1$. We fit using only ZTF observations and by combining ATLAS, GOTO, and ZTF data, resulting in a total of four BayeSN fits. The resulting fits are shown in Fig.~\ref{BayeSN_fits}. All fits result in the same E$($B$-$V$)_\text{host}$ within errors, with marginally lower values when $R_V=3.1$. BayeSN does not take into account that the MW $R_V$ can vary between different parts of the sky. To account for this systematic uncertainty we add 0.03 mag to the \ebv$_{host}$ error in quadrature, as was done for the Lira law. Fitting the three surveys together consistently reduces the extinction estimate and uncertainty. We compared our results to a fit that also included the LT data but found no change in the $A_V$ estimates. The ZTF+ATLAS+GOTO BayeSN E$($B$-$V$)_\text{host}$ value of $0.19\pm0.04$ mag is consistent with that obtained from the SALT2 of $c=0.2105\pm0.0008$ mag within the uncertainties, and suggests that the observed colour is mainly due to host extinction while intrinsic value of $c\sim0$ for SN~2024gy.

\subsection{Spectroscopic methods}
\label{host_ext_spec}
\paragraph{Spectral matching.}
The host extinction can also be estimated by matching spectra to a similar SN with low or known extinction \citep[e.g.][]{SM_2008J, Cikota16, 2012cu} at similar epochs. We do this by matching the spectra to those of SN~2011fe \citep[\ebv$_{host}=0.014\pm0.002$~mag,][]{Patat_11fe} and SN~2017erp \citep[\ebv$_{host}=0.11\pm0.03$~mag,][]{17erp_Brown}. The comparison spectra have been corrected for both MW and host galaxy extinction. After correcting the SN~2024gy spectra for MW extinction, we estimate the host extinction by fitting the \citet{F99} dust law in such a way that the spectra match the comparison spectra, assuming $R_V=3.1$. As emission and absorption lines can have different strengths in different objects, we focus on comparing regions that primarily show the continuum (e.g. 6800--7200~\AA\ in the early and peak spectra, and the regions around 4500--4600~\AA\ and on either side of the \SII\ feature in the peak spectrum). The bluer regions are more sensitive to the amount of removed extinction while the redder regions help with scaling the spectra. We therefore limit our comparison to spectra taken at phases after $-10$ days as the earliest spectra look anomalous due to e.g. the HV \CaII\ NIR component, and before $+25$ days to avoid going into the nebular phase where large amounts of \FeII\ and \FeIII\ lines dominate the spectra. 

Table~\ref{specmatch_table} lists the phases at which the comparisons spectra were taken and the host extinction that best matched SN~2024gy to these spectra. The best matches with SN~2011fe have an \ebv$_{host}=0.24 \pm 0.02$ mag when assuming $R_V=3.1$. The best matches with SN~2017erp have a wider extinction spread and on average give a slightly lower \ebv$_{host}=0.19 \pm 0.04$ mag. Using $R_V = 2.659$ \citep[the default value for the BayeSN model,][]{BayeSN_W22} instead changes the required $A_V$ but gives the same \ebv\ within the uncertainty. After comparing both $R_V$ values we do not significantly favour one over the other. The difference in the found \ebv$_{host}$ when using different comparison objects shows the limitation of this method, and extinction in the comparison object will add additional uncertainty. We therefore choose the result from the SN~2011fe matching as this SN had a significantly lower \ebv$_{host}$, though this does increase our extinction estimate for this method.

\paragraph{ISM absorption features.}
\label{ISM_features}
\begin{figure*}
    \centering
    \includegraphics[width=0.97\textwidth]{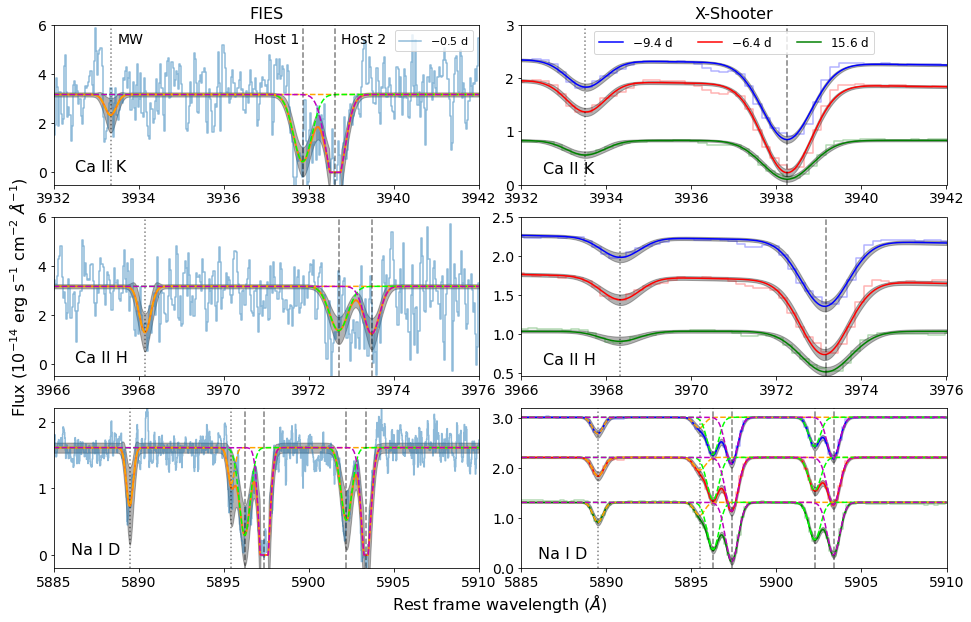}
    \caption{The \CaII\ H\&K and \NaI\ D regions in the $-$ 5 d FIES and in the three X-Shooter spectra. The pre-peak X-Shooter spectra have a vertical offsets for readability purposes. The fitted model is shown as a solid line with $3\sigma$ uncertainties shown as a gray band. The FIES model consists of three sets (one MW, two host) of four Gaussian functions (two for the \CaII\ H\&K doublet and two for the \NaI\ D doublet) to model the absorptions lines which are shown with dashed lines. The centroids are redshifted by a common factor $z$ for each set of absorption lines. The MW and host components are marked with dotted and dashed vertical lines, respectively. Due to the lower resolution of the X-Shooter spectra, the absorption lines are blended and cannot be fitted separately for the host \CaII\ H\&K, and the common $z$ requirement between the \CaII\ and \NaI\ D regions is removed.}
    \label{Ca_Na_abs}
\end{figure*}

\begin{figure*}
    \centering
    \includegraphics[width=0.97\textwidth]{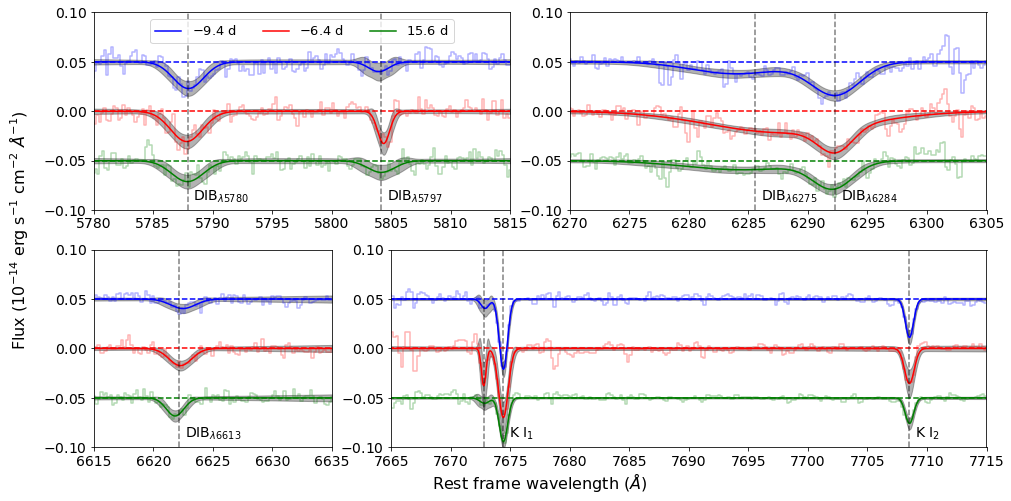}
    \caption{The continuum subtracted \KI\ and DIB regions in the three X-Shooter spectra are shown in the paler histograms. The fitted models are shown as solid lines with $3\sigma$ uncertainties as gray bands on each model fit. Each feature is labelled with a dashed vertical line. Only for the \KI$_1$ line we recovered multiple components at different velocities.}
    \label{K_DIBs_abs}
\end{figure*}

Both the \NaI\ D and \CaII\ H\&K doublets are well-resolved in SN~2024gy in the high-resolution FIES spectrum (R $\sim67,000$) at $-5$ days with respect to peak, as well as in the three X-Shooter spectra. The FIES spectrum is shown in the left panels of Fig.~\ref{Ca_Na_abs} and three absorption components for each line are marked, with most of them partially overlapping each other. These components correspond to three clouds (one MW component, two host components) of absorbing material. Using \textsc{lmfit}\footnote{\url{https://lmfit.github.io/lmfit-py/index.html}} \citep{lmfit}, we fit three sets of four Gaussian absorption lines (one set for each cloud) redshifted by a common factor on top of a background. We fit the background as a constant value in the fitted regions, but allow the value to change between the \CaII\ H\&K and \NaI\ D regions. The absorption lines are flat-topped if the line saturates \citep[e.g.][]{Phillips_Na_overpredict_ebv}. As the width of an absorption line is linked to the velocity distribution in the ISM cloud, we link the line widths as such. We assume that the clouds are well mixed, so we use one width for both the \CaII\ H\&K and \NaI\ D regions per cloud.As absorption profiles are not always best modelled by a Gaussian profile \citep[see e.g.][]{voigt_approx} we also fitted Lorentzian and Voigt profiles, but find the best fits when using Gaussian profiles.

The result is a fit of two continua and 12 absorption lines to model the \CaII\ H\&K and \NaI\ D regions, simultaneously. The resulting fit is shown for the FIES spectrum on the left side of Fig.~\ref{Ca_Na_abs}. The first cloud has $z < 0$, meaning that this is a MW cloud at $v\approx -25\pm1$ km s$^{-1}$. The other two clouds have redshifts similar to that of SN~2024gy, showing that these clouds are in NGC~4216 but are offset with respect to the SN. One cloud is blueshifted by $v=24\pm1$ km s$^{-1}$ compared to the SN, while the other is redshifted by $v=35\pm1$ km s$^{-1}$. The equivalent width (EW) of each line is shown in Table~\ref{EWs}, and we use the relations from \citet{NaID_EBV_relation} to convert the \NaI\ D EW into \ebv\ estimates for each cloud.

The X-Shooter spectra have lower resolution (R $\sim5400-8900$), resulting in more blended absorption lines (right panel of Fig.~\ref{Ca_Na_abs}). As a result, the \CaII\ H\&K absorption in NGC~4216 is fit with a single Gaussian at the mean redshift, though the \NaI\ D region still has two distinct host components. We fit the \CaII\ and \NaI\ D regions separately for each of the three X-Shooter spectra at $-9.4$, $-6.4$ and $+15.6$ days with respect to peak and show the fits in Fig.~\ref{Ca_Na_abs}. We convert the \NaI\ D EWs to \ebv\ estimates using the relations of \citet{NaID_EBV_relation} (Table~\ref{EWs}).

Several SNe have shown time variability in their narrow absorption lines, indicating that the absorbing region is CSM close to the explosion site that is reacting to the intense SN radiation (e.g. \NaI\ D in SN~1999cl; \citealt{1999cl_NaID} and SN~2006X; \citealt{Patat_06X}, and \KI\ in SN~2014J; \citealt{14J_time_varying_K}). The measured \NaI\ D EWs in SN~2024gy are consistent within their uncertainties across the different epochs, suggesting that the line-forming clouds are more likely further from the explosion site.

We find several absorption features in the X-Shooter spectra that line up with the \KI\ doublet and several DIBs. For the \KI\ doublet we fit Gaussian absorption features as well as a linear continuum, which is shown after continuum removal in the bottom right panel of Fig.~\ref{K_DIBs_abs}. We recover both components of the doublet for the higher redshift cloud and tentatively recover \KI$_1$ 7665 \AA\ in the other host galaxy cloud. This latter line is at the edge of a telluric absorption feature, making it very sensitive to the exact correction applied by \textsc{molecfit}. The first absorption region shows that \KI$_1$ is somewhat brighter than \KI$_2$, so the second absorption region of \KI$_2$ is likely too faint given how weak the second absorption of \KI$_1$ is. No MW absorption is seen. 

Figure~\ref{K_DIBs_abs} also shows the recovered DIBs at 5780, 5797, 6275, 6284 and 6613 \AA. These regions have more complex background shapes, so we fit them with a second order polynomials where required. \citet{APO_DIB_cat} presented a large catalogue with 559 identified DIBs. They show that the DIBs we find likely contain contributions from different DIBs located close together, adding uncertainties on their measurements. However, assuming the wavelength of the strongest DIB in the region, we find that all recovered DIBs have redshift values that correspond to the higher redshift cloud in the \NaI\ D, \CaII\ H\&K, and \KI\ lines. This is also consistently the strongest component for each atomic absorption line considered above, showing that this cloud has the highest column density within the line-of-sight. We use these fits and the relations from \citet{KI_EBV_relation} and \citet{DIB_EBV_relation} to estimate the \ebv, which are listed in Table~\ref{EWs}.

To get a single \ebv$_{host}$ estimate for each absorption line, we sum over the host galaxy components (if we measured multiple) and take the weighted mean for each line. The resulting values are shown in Table~\ref{reddening_estimates}. Finally, we take the mean value over all lines to find the mean extinction estimate using the absorption lines, giving \ebv$_{host}=0.12\pm0.02$ mag. The \NaI\ D estimate is excluded in this final mean as it is clearly an outlier due to one of the components saturating (see Section~\ref{ISM_lines_discussion}).

\subsection{Polarimetry method}
\label{host_ext_pol}
Elongated dust grains absorb the most light along their major axis. When these grains are aligned by galactic magnetic fields, there will be a preferential direction in which light will be absorbed, which can induce a polarisation \citep{Davis_1951, Spitzer_51, Davis_1959}. \citet{Serkowski_ISM_pol} showed that while there is a large variation in the amount of polarisation that ISM can cause, there is a well-defined upper limit of $P_{max}(\%) = 9\times\ebv$. These measurements were made for MW ISM and therefore assume $R_V=3.1$. This limit is often used to estimate the maximum contribution of ISM in a strong polarisation signal, but assuming that this relation also holds for other galaxies, we can use it the other way around as
\begin{equation}
    \text{\ebv} = \frac{P_{max}(\%)}{9} \geq \frac{P_{obs}(\%)}{9}.
    \label{pol_eqn}
\end{equation}

From this we find that an \ebv$\geq 0.08\pm0.01$ can explain the measured polarisation as being entirely interstellar polarisation (ISP). This estimate is lower than any of our previous estimates, which is consistent with the expectation that the SN does not have an intrinsic continuum polarisation. When we investigated the narrow ISM absorption features in Sect.\ref{ISM_features}, we found evidence of multiple clouds in the host galaxy in front of the SN. This can explain the observed angle dependency if we assume the clouds have different properties and grain orientations \citep{Serkowski_ISM_pol}.

\subsection{Summary}
We presented five methods to estimate host extinction. Each method has a similar uncertainty in their final \ebv$_{host}$ estimate and they differ by up to $3\sigma$. The Lira law, BayeSN, and spectral matching methods give very similar \ebv$_{host}$ estimates. We take the mean of these three methods (assuming $R_V=3.1$ and using the BayeSN result with all data) to find \ebv$_{host}=0.22\pm0.04$ mag across these methods. Of the five extinction estimates using ISM lines, two give similar values as the other three methods but have a large uncertainty, two give lower extinction estimates with much smaller uncertainties, and one is invalid as the absorption line is saturated. The result is a lower mean \ebv$_{host}$ estimate found through ISM lines.

\section{Discussion}
\label{discussion}
The location of SN~2024gy was observed by multiple surveys, providing long baseline, well-sampled multi-colour optical light curves. Multi-epoch optical and NIR spectroscopy were also obtained from $-$15.2 days to 316.8 days with respect to peak. Both photometrically and spectroscopically, SN~2024gy looks like a normal SN Ia, reddened by intervening dust. We find no signs of precursor activity brighter than $M=-8.75$~mag (assuming no host extinction) in the last 9.5 years before its explosion or an enhanced late-time signal on top of the declining SN tail. The main irregularity is the presence of a persistent strong HV \CaII\ NIR feature at pre-peak phases with velocities over 30\,000~km~s$^{-1}$, which matches better with HV SNe such as SN~2019ein.

\subsection{\ebv\ estimates through ISM absorption features}
\label{ISM_lines_discussion}
Theoretically, for a gas that traces the extinction with a suitable absorption feature the relation $\text{EW} \propto A_V$ should hold at low column densities. The relation will also depend on factors such as the ratio between the gas and dust and the absorption coefficient for the feature, and different clouds can have different dust properties, affecting the proportionality constant between EW and $A_V$ \citep[e.g.,][]{NaID_EBV_relation}. Therefore, it is only possible to find an empirical relation and assume that the proportionality does not fluctuate too much between clouds. 

Extensive studies have been performed to find relations between EW and extinction, but most have their roots in MW studies \citep[e.g. by measuring the extinction to certain stars or by using extragalactic objects but focusing on the MW absorption;][]{Ca_II_bad_dust_estimator, KI_EBV_relation, NaID_EBV_relation}. These relations intrinsically assume that the ISM in the MW and host galaxies are similar enough for these relations to still hold. Moreover, the theoretical relation is based on a proportionality between EW and $A_V$, but they are usually presented as being between EW and \ebv. This means a value for $R_V$ has been assumed, usually $R_V=3.1$ as the focus lies on MW dust. But if $R_V \neq 3.1$ in other galaxies, as numerous studies suggest \citep[e.g.][]{14J_odd_Rv, SUGAR_lc_fitter, Rose_22_Rv, BayeSN_W22}, this adds uncertainty in the host \ebv\ values.

The relations assume low column densities but they break down if the line saturates, as is the case for \NaI\ D in SN~2024gy. Whether or not the line is saturated is however not always clear, as \NaI\ D does not appear to saturate in the X-Shooter spectra. This shows that even if strong lines do not appear to fully saturate, the estimated \ebv\ values might be overestimated. Some studies also showed that some SNe seem to follow a different \ebv\ to \NaI\ D relation compared to others \citep[see e.g.][and references therein]{Turatto_2003_proceedings}. \citet{Phillips_Na_overpredict_ebv} show objects with strong \NaI\ D absorption have multiple absorption features with the bluer components being weaker \citep[blueshifted in the classification scheme of][]{Sternberg11}. SN~2024gy falls in the same category, and similar double profiles are found in the \CaII\ H and K lines, and in \KI $\lambda7665$. While \citet{Phillips_Na_overpredict_ebv} try to associate blueshifted \NaI\ D components to outflowing CSM, they do note objects with confirmed CSM generally do not show this \NaI\ D feature.

Weak lines such as DIBs are also difficult to measure accurately, as they can appear to evolve as the slope of the background continuum changes \citep{Narrow_absorbers_in_SN_spectra}. The variations we find in the DIB absorption in SN~2024gy are consistent with this, and are not interpreted as changes in the absorbing material. These effects add uncertainty to the measured values. However, they are often smaller than the dispersion in the relations themselves, which limits the precision at which these relations can be used. The inverted argument can still be used though: the absence of narrow absorption lines, especially in high resolution and high S/N spectra, can show that the amount of intervening dust is below some upper limit. If narrow lines are detected, it is best to measure them in multiple spectra, take mean \ebv\ estimates for each line, and combine to get a weighted mean \ebv\ value from all useable ISM lines.

The MW \NaI\ D component is much weaker and far from saturating. The combined mean extinction value for this component is \ebv$_{MW}=0.043$ mag, which is higher than the value given by the \citet{SFD98_dust_maps} dust maps (\ebv$_{MW}=0.023$ mag). The scatter on the \NaI\ D relation is 0.008 mag \citep{NaID_EBV_relation} and the formal uncertainty on the dust maps is $10\%$ \citep{SFD98_dust_maps}, which puts these values just below $2.5\sigma$ from each other.

\subsection{\ebv\ estimates through different methods}
The Lira law method performs quite well, but requires the light curve to be well observed in several filters for several months to get a decent estimate. At least two filters are required to estimate \ebv$_{host}$, and three are required to disentangle its degeneracy with $R_V$. Distant SNe that only rise above a survey's detection limit while they are around their peak cannot be detected at late phases, preventing the use of this method. In addition, to avoid adding systematic uncertainties a reddening-free sample observed in the same filters is needed to compare against. Here we used a sample of ZTF objects as both SN~2024gy and the comparison sample have been observed with the exact same instrument and setup.

The Lira law is thus a powerful tool if the requirements can be met, but it comes with a few assumptions that can be easy to overlook. First of all, a reddening free sample is free from extinction in the host galaxy and the MW. Simply selecting objects at the outskirts of their host galaxies to reduce the amount of host galaxy dust in the line-of-sight is not enough, as the MW can still add a significant foreground of foreground extinction. We minimize this effect in Sect.~\ref{host_ext_phot} by requiring \ebv$_{MW}<0.1$ mag and correct for the remaining extinction assuming $R_V=3.1$. The assumed MW $R_V$ value also carries an uncertainty, as it is shown to vary across the sky \citep{MW_Rv_map}. More than likely it differs slightly in each direction of interest. However, it is often not practical to obtain a separate MW $R_V$ estimate for each object. An \ebv$_{MW}$ cut limits the effect an error in the MW $R_V$ can have, but it has to be taken into account. Finally, in order to be able to compare the colours of different objects, K-corrections must be applied to account for the changing effective probed wavelength region at various redshifts. As not every data point will have an associated observed spectrum to calculate the exact K-correction to apply, one must instead assume that different objects look similar enough to interpolate using their spectra instead. While this assumption is reasonable for a properly chosen sample of objects, there is some amount of uncertainty associated with it that should be propagated.

When fitting the photometry with BayeSN, we fitted with only ZTF observations and also by combining observations from ZTF, ATLAS, and GOTO. Adding more data lowered the $A_V$ estimates by about $1\sigma$. The default value of $R_V=2.659$ is the best fit value found by \citet{BayeSN_W22} when training BayeSN. Assuming $R_V=3.1$ results in only small changes to the \ebv\ values ($0.5\sigma$ lower). \citet{Kwok_24gy} estimated the host extinction for SN~2024gy using BayeSN, a different light curve dataset, and assumed $R_V=2.55$ to find a value of \ebv$_{host}=0.28\pm0.03$ mag , consistent within $2\sigma$ of our BayeSN estimate of $0.19\pm0.04$ mag (assuming $R_V=3.1$) for our full light curve fit. 

The spectral matching technique is dependent on the chosen wavelength regions to compare, which is phase-dependent. Suitable regions are dominated by the continuum and lack spectral features which may differ in strength between SNe. Difference in phase and brightness between the spectra being compared can add additional uncertainty through the need of flux rescaling. Despite this, the method finds a \ebv\ value close to the results from the photometric methods. However, this method is not sensitive enough to favour a particular $R_V$ value.

Polarimetric observations are not commonly used to estimate extinction as multiple high S/N observations are required. In contrast, high-cadence photometry is taken by all-sky surveys, providing a much larger dataset that is readily available without manual scheduling. Also, the reduction process is more complicated for polarimetry, whereas survey data is generally provided in the reduced state. Finally, the relation between polarisation degree and \ebv\ is an upper limit on the amount of polarisation a certain amount of dust can induce with a wide spread below this limit \citep{Serkowski_ISM_pol}. Unsurprisingly, inverting the relation to estimate \ebv\ instead leads to a lower limit that is not very constraining compared to other methods.

Finally, the Balmer decrement method using the host galaxy spectrum is much higher than any of the estimates using the SN observations. As noted before, the SDSS spectrum of the host was taken about 5\arcsec from the SN location, which might be in a more dusty part of the galaxy. We were unable to find a galaxy trace in the 2D SN spectra to use the Balmer decrement method at a location closer to the SN. The method as described in \citet{Balmer_decrement_ebv} assumes the galaxy is star-forming, but radio observations taken by \citet{Virgo_radio} show that NGC~4216 is an anemic galaxy as it has a very low \HI\ column density and star-formation rate. This could suggest that the Balmer decrement method may not be applicable in the same way for this type of galaxy.

Several of our methods implicitly assume $R_V=3.1$, while others are insensitive to changes in the assumed $R_V$. When \citet{Li_24gy} assume $R_V=3.1$ the resulting \ebv$_{host}$ is within $1\sigma$ of our BayeSN results and consistent with \citet{Kwok_24gy}. While they do detect host \NaI\ D in their spectra, they do not use this to estimate the host extinction. 

\citet{Li_24gy} also use several other methods to estimate host extinction. First, they use the Lira law and a method incorporating $\Delta m_{15}$ and the $($B$-$V$)$ colour at maximum light as presented in \citet{Phillips_rel2} to estimate \ebv$_{host}=0.40\pm0.03$. Then they use SUGAR \citep{SUGAR_lc_fitter} to obtain $A_V=0.662 \pm 0.018$ mag, and combine this with their previous \ebv$_{host}$ result to get $R_V\sim1.5$. Finally, they use this $R_V$ value in their MLCS2k2 \citep{MLCS2k2_lc_fitter} and use the resulting \ebv$_{host}$ throughout their analysis. However, \citet{SUGAR_lc_fitter} reported that SUGAR fits best with a \citet{CCM89} extinction law with $R_V=2.6$ which would give \ebv$=0.255 \pm 0.007$ mag, consistent with \citet{Kwok_24gy} and our results.

\subsection{Explosion models}
We compared observed features of SN~2024gy to predictions for the double-detonation and delayed-detonation models using several methods. First, we measured a Ni over Fe abundance ratio of $0.11\pm0.05$ at 316.8 days post peak for SN~2024gy. This is consistent with delayed-detonations of M$_\text{Ch}$ WDs \citep{Seitenzahl_13, Maguire18}. This value is also marginally consistent with the range predicted by double-detonations of sub-M$_\text{Ch}$ WDs \citep{shen18} but requires solar metallicity. In Sect.~\ref{tardis}, we also compared the spectra at -8.5 d and -5.5 d to the \citet{GRONOWDDMODELS} double-detonation models and the N100 delayed-detonation model from \citet{N100MODEL} using \textsc{tardis}. While both types of models matched reasonably well with the observations, the double-detonation consistently reproduced the observed HV component in \CaII\ NIR while it was absent in the N100 model (although see section~\ref{line_vel_disc}). The best double-detonation matches were viewed from the pole at which the \ion{He}{} occurs on a WD with a $0.9~M_\odot$ core and a $0.05~M_\odot \lesssim M_{He} \lesssim 0.1~M_\odot$ \ion{He}{} shell. Our third method, which uses the blue wing edge of the $1.57\ \mu$m feature to measure the outer velocity of $^{56}$\ion{Ni}{}, should theoretically be able to differentiate between $M_{Ch}$ and sub-$M_{Ch}$ mass models. However, as we do not have the models to compare against, the result remains inconclusive. Nevertheless this method is a promising additional test if developed further. 

By binning the late-time observations we were able to detect the SN tail up to over 400 days after the peak, and found no signs of a flux excess that could indicate interaction with CSM. Similarly, by binning the pre-SN observations we searched for possible precursor events but found none brighter than $M\approx-8.75$ mag between 2015 and 2024. These findings are in line with the observations that SNe Ia typically explode in rather clean environments \citep{Margutti_11fe_Xray, Margutti_14J_Xray, Maguire_H_uplims, Lundqvist_no_radio_Ia, 17cbv_20nlb_CSM_constraints}, and events with signs of CSM are rare \citep[e.g.][]{Sharma_Ia_CSM, JHT_DR2}. While a clean environment does not rule out specific explosion models it disfavours progenitor models that can pollute the progenitor environment, such as a CSM created by the companion stellar wind \citep{Hayden_2010, Lundqvist_2015} in single degenerate systems or repeated stripping of the secondary WD over multiple orbits in double degenerate systems with high initial eccentricity \citep[e.g.][]{2020aeuh_Tsalapatas}.

SN~2024gy shows some features that favour double-detonation scenario, while other features favour a delayed-detonation scenario. We therefore cannot favour one over the other based on our results. A further exploration of this topic is however outside the scope of this work. As discussed in Sect.~\ref{intro}, \citet{Li_24gy} favoured a M$_\text{Ch}$ delayed-detonation model for SN~2024gy based on the presence of HV features in the \CaII\ NIR triplet, strong lines of \SiII, and the high Ni/Fe ratio measured in their nebular phase spectra. \cite{Kwok_24gy} relate profile shapes to the distribution of each element and find signs of enhanced central Ni production. They also favoured a M$_\text{Ch}$ delayed-detonation explosion model, as it reproduces the observed emission line profiles in their JWST spectra.

\subsection{Line velocity evolution and high velocity features}
\label{line_vel_disc}
We find a HV component in \SiII~$\lambda6355$ more than 10 days before peak brightness.  The HV \CaII\ NIR feature is persistent, being detected up to near peak brightness. This feature is also remarkably strong and wide at early epochs, with some of the highest velocities recorded to date even when comparing to other objects that are known for their HV features. \citet{Li_24gy} find similar extreme values for the HV \CaII\ NIR triplet, which they try to explain using different mechanisms. One of the discussed mechanisms is early interaction with a relatively dense cloud or shell of CSM, causing density enhancements and ionization effects. They point out that \NaI\ D can show evidence of CSM if a blue-shifted component is found that evolves with the SN, and calculate that such a component should have a EW~$\approx0.5$~\AA\ and vary by $\approx0.1$~\AA\ if the \NaI\ D absorption is significantly affected by CSM interaction. They find no evidence of evolving narrow lines, though the resolution of their spectra prevented them from detecting components smaller that EW~$\sim3$ \AA.

With our high resolution spectra we do find multiple host components in the \NaI\ D lines of SN~2024gy, with the bluest component having a EW~$\approx0.5$~\AA. However, we do not see significant evolution in either of the host components, disfavouring the mechanism suggested by \cite{Li_24gy}. Moreover, we find the same multi-component absorption in \CaII\ H\&K, and \KI~$\lambda7665$, which also do not show significant signs of evolution. \NaI\ D variability can be quite subtle \citep{Patat_06X, Sternberg11}, and small variability could be too subtle for to be detected with X-Shooter spectra.

Other explanations for the HV \CaII\ NIR triplet include abundance enhancements in a delayed-detonation or double-detonation model. In a delayed detonation, outward mixing in the outer layers could explain the HV features, as has been proposed in e.g. SN~2019ein \citep{2019ein_pellegrino}. However, unlike SN~2019ein, other lines in SN~2024gy do not show HV components as strong or persistent as the \CaII\ NIR triplet, the only other detected HV component faded fast. This would suggest a different density structure in SN~2024gy.

The double-detonation scenario is inherently asymmetric, with large differences in ejecta structure along different viewing angles. The explosion of a helium shell creates the required abundance enhancements, which can then be pushed to the HV regime along certain viewing angles. However, our early SN~2024gy detections are too sparse in any given band to rule out the presence of a small bump. \citet{Li_24gy} have a better coverage of the early light curve, but do not report any evidence of an early excess either.

\section{Summary and Conclusions}
\label{conclusion}
Using five surveys and over a dozen different telescopes we collected a rich, heterogeneous optical photometric and spectroscopic dataset of SN~2024y, as well as optical polarimetric data and NIR spectroscopy. Initial light-curve fitting using \textsc{salt2} showed that while SN~2024gy has relatively normal parameters, its colour $c$ indicated a some amount of extinction. We then used our rich dataset to obtain host extinction estimates using different techniques and compared them to one another. Our main conclusions are listed below:

\begin{enumerate}
    \item There is a significant amount of dust in front of SN~2024gy. By combining different extinction estimators, we find a mean host extinction of \ebv$_{host}=0.22\pm0.04$ mag , assuming $R_V=3.1$. While three of the methods give similar results, the ISM features have a larger spread when different lines with estimates ranging between $0.1\lesssim$ \ebv$_{host}\lesssim0.2$ mag. These estimates are somewhat lower than the values found by \citet{Li_24gy} and \citet{Kwok_24gy}.
    \item All our methods are either insensitive to changes in $R_V$ or implicitly assume $R_V=3.1$ as they are calibrated using MW sources. We therefore cannot make a claim on whether the dust in NGC~4216 and MW dust have different $R_V$ values.
    \item \NaI\ D is inadequate for \ebv$_{host}$ estimation in SN~2024gy as the FIES spectrum shows that it has saturated components. In lower resolution spectra this is not obvious, which could lead to a false estimate for the host extinction. Generally, it is easier to say that the absence of \NaI\ D absorption suggests negligible extinction than to estimate the amount of extinction when \NaI\ D is present.
    \item The other narrow lines give better extinction estimates, though some still suffer from large uncertainties. We attempted to mediate this by taking the weighted mean over all the measurements. However, as a general method for other SNe this may be difficult, as most ISM lines are weak and cannot be used to estimate extinction unless high resolution data is available. Most of the relations are also derived empirically from MW dust measurements, intrinsically assuming a value for $R_V$ that may be incorrect for other galaxies.
    \item The observed polarisation of SN~2024gy is likely primarily induced by the host ISM. By using a relation between the extinction and maximum amount of polarisation that can be induced by the ISM, we found a consistent but not constraining lower limit on the amount of host extinction. The relative difficulty of obtaining polarimetry data combined with the nature of this extinction estimation method makes this method impractical to estimate extinction in SNe in general.
\end{enumerate}

\begin{acknowledgements}
      J.H.T., K.M., T.M.B., and U.B. acknowledge Horizon Europe ERC grant no. 101125877. C.O.D acknowledges funding from the Irish Research Council (IRC; GOIPG/2024/4378), M.P. from a UK Research and Innovation Future Leaders Fellowship (MR/020784/1, UKRI1062). B.G. acknowledges the UKRI’s STFC studentship grant funding (ST/X508871/1). L.H. acknowledges the IRC (GOIPG/2020/1387). C.L. is supported by DoE award \#\,DE-SC0025599. W. M. Keck Observatory and MMT Observatory access was supported by Northwestern University and the Center for Interdisciplinary Exploration and Research in Astrophysics (CIERA). A.A. is funded by financial support from AGAUR, CSIC, MCIN and AEI 10.13039/501100011033 under projects PID2023-151307NB-I00, PIE 20215AT016, and CEX2020-001058-M. T.-W.C. acknowledges support from the Ministry of Education Yushan Fellow Program (MOE-111-YSFMS-0008-001-P1) and from the National Science and Technology Council, Taiwan (NSTC 114-2112-M-008-021-MY3). J.D. thanks the Fundação para a Ciência e Tecnologia (FCT), Portugal (PhD grant 2023.01333.BD) and the Center for Astrophysics and Gravitation (CENTRA/IST/ULisboa) through grant No. UID/PRR/00099/2025 and grant No. UID/00099/2025. L.G. acknowledges financial support from AGAUR, CSIC, MCIN and AEI 10.13039/501100011033 under projects PID2023-151307NB-I00, PIE 20215AT016, and CEX2020-001058-M. Y.-L.K. was supported by the Lee Wonchul Fellowship, funded through the BK21 Fostering Outstanding Universities for Research Program (4120200513819) and the National Research Foundation of Korea to the Center for Galaxy Evolution Research (RS-2022-NR070872, RS-2022-NR070525, RS-2026-25473561). R.S. thanks the Fundação para a Ciência e Tecnologia (FCT), Portugal, (PhD grant 2024.03599.BD) and to CENTRA/IST/ULisboa (UID/PRR/00099/2025,  UID/00099/2025).
      
      Based on observations obtained with the Samuel Oschin Telescope 48-inch and the 60-inch Telescope at the Palomar Observatory as part of the Zwicky Transient Facility project. ZTF is supported by the National Science Foundation under Grants No. AST-1440341 and AST-2034437 and a collaboration including current partners Caltech, IPAC, the Oskar Klein Center at Stockholm University, the University of Maryland, University of California, Berkeley, the University of Wisconsin at Milwaukee, University of Warwick, Ruhr University Bochum, Cornell University, Northwestern University and Drexel University. Operations are conducted by COO, IPAC, and UW. 

      SED Machine is based upon work supported by the National Science Foundation under Grant No. 1106171
      
      The Gordon and Betty Moore Foundation, through both the Data-Driven Investigator Program and a dedicated grant, provided critical funding for SkyPortal.
      
      The ztfquery code was funded by the European Research Council (ERC) under the European Union's Horizon 2020 research and innovation programme (grant agreement n°759194 - USNAC, PI: Rigault).
      
      ATLAS is primarily funded through NASA grants NN12AR55G, 80NSSC18K0284, and 80NSSC18K1575. The ATLAS science products are provided by the University of Hawaii, Queen's University Belfast, STScI, SAAO, and Millennium Institute of Astrophysics in Chile.

      GOTO acknowledges the Monash-Warwick Alliance; University of Warwick; Monash University; University of Sheffield; University of Leicester; Armagh Observatory \& Planetarium; the National Astronomical Research Institute of Thailand (NARIT); Instituto de Astrofísica de Canarias (IAC); University of Portsmouth; University of Turku; University of Birmingham; and the UK STFC (ST/T007184/1, ST/T003103/1 and ST/Z000165/1).
      
      Based on observations made with the NOT, owned in collaboration by the University of Turku and Aarhus University, and operated jointly by Aarhus University, the University of Turku and the University of Oslo, representing Denmark, Finland and Norway, the University of Iceland and Stockholm University at the Observatorio del Roque de los Muchachos, La Palma, Spain, of the Instituto de Astrofisica de Canarias.
      
      The data presented here were obtained [in part] with ALFOSC, which is provided by the Instituto de Astrofisica de Andalucia (IAA) under a joint agreement with the University of Copenhagen and NOT.

      The INT is operated on the island of La Palma by the Isaac Newton Group of Telescopes in the Spanish Observatorio del Roque de los Muchachos of the Instituto de Astrofísica de Canarias.
      
      The LT is operated by Liverpool John Moores University  with financial support from the UK STFC.
      
      This work makes use of observations from the Las Cumbres Observatory global telescope network.

      Some of the data presented herein were obtained at Keck Observatory, which is a private 501(c)3 non-profit organization operated as a scientific partnership among the California Institute of Technology, the University of California, and the National Aeronautics and Space Administration. The Observatory was made possible by the generous financial support of the W. M. Keck Foundation. The authors wish to recognize and acknowledge the very significant cultural role and reverence that the summit of Maunakea has always had within the Native Hawaiian community. We are most fortunate to have the opportunity to conduct observations from this mountain.
      
      Observations were obtained at the MMT Observatory, a joint facility of the Smithsonian Institution and the University of Arizona.

      Based on observations collected at the European Organisation for Astronomical Research in the Southern Hemisphere, Chile, as part of ePESSTO+  (ID 112.25JQ). 

      This work made use of HESMA (https://hesma.h-its.org). The development of \textsc{tardis} received support from GitHub, the Google Summer of Code initiative, and from ESA's Summer of Code in Space program. \textsc{tardis} is a fiscally sponsored project of NumFOCUS. \textsc{tardis} makes extensive use of Astropy and Pyne.

      This research made use of \textsc{tardis}, a community-developed software package for spectral synthesis in supernovae \citep{2014MNRAS.440..387K, kerzendorf_2025_18072609}. 
      
\end{acknowledgements}

\section*{Data Availability}
The photometry and polarimetry data used in this paper is available in electronic form at the CDS via anonymous ftp to \url{cdsarc.u-strasbg.fr} (130.79.128.5) or via \url{http://cdsweb.u-strasbg.fr/cgi-bin/qcat?J/A+A/}. The spectra are available on WISeREP (\url{https://www.wiserep.org}). All data is also available upon request to the author.

\bibliographystyle{aa}
\bibliography{refs}

\begin{appendix}
\onecolumn
\section{Figures}
\begin{figure*}[hp!]
    \centering
    \includegraphics[width=\textwidth]{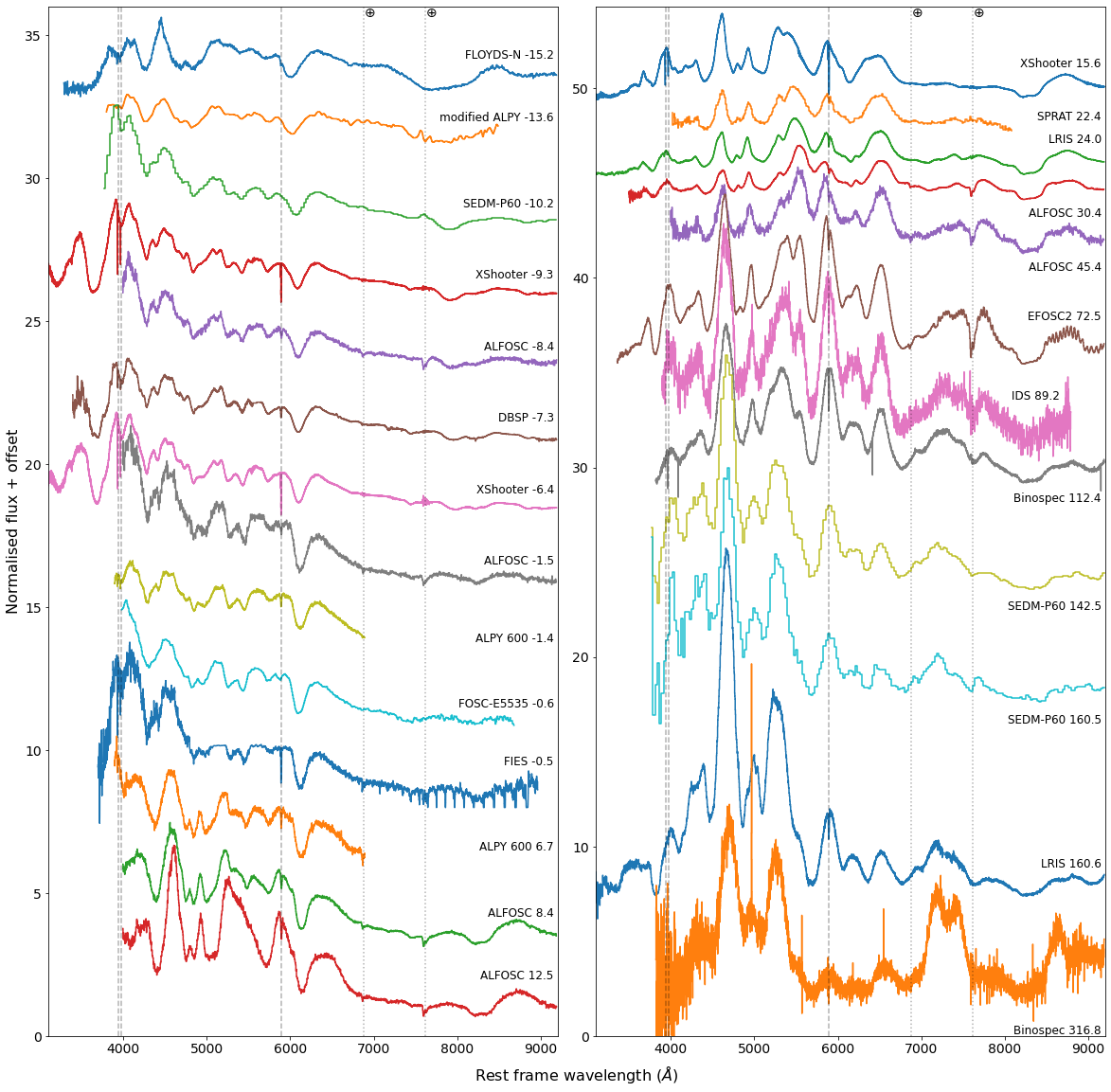}
    \caption{Optical spectra of SN~2024gy in the SN rest frame. All spectra are corrected for MW extinction and normalised to the mean flux in the region between 6760~\AA\ and 6830~\AA. The FIES spectrum has been rebinned for clarity. The dashed vertical lines mark the locations of the \NaI\ D and \CaII\ H\&K lines, and the dotted lines mark the location of the largest telluric regions.}
    \label{all_optical_specs}
\end{figure*}

\begin{figure*}[h!]
    \centering
    \includegraphics[width=0.6\textwidth]{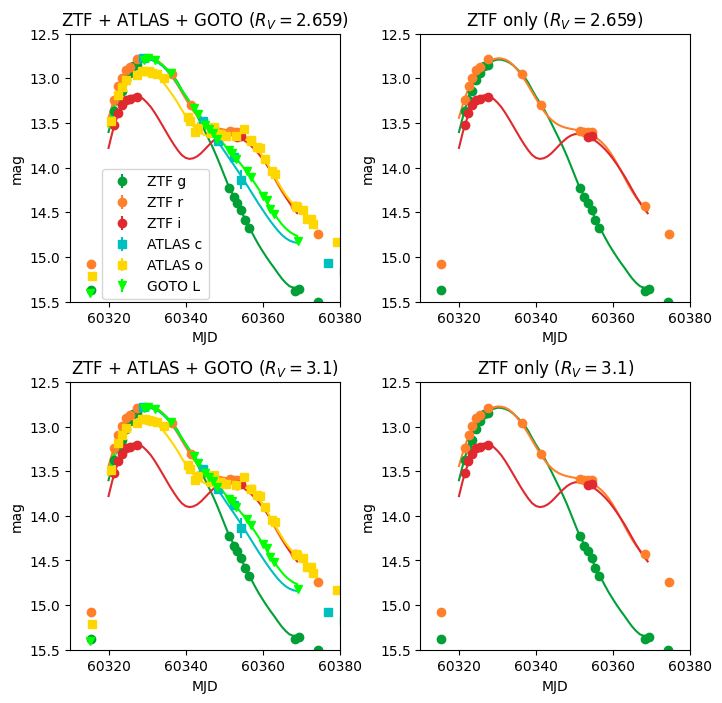}
    \caption{BayeSN fits of SN~2024gy using the \citet{BayeSN_W22} model. The top row assumes the default $R_V$ while the bottom row assumes a MW $R_V$. In the left column we use all data, on the right we only use ZTF data.}
    \label{BayeSN_fits}
\end{figure*}

\begin{figure*}[h!]
    \centering
    \includegraphics[width=\textwidth]{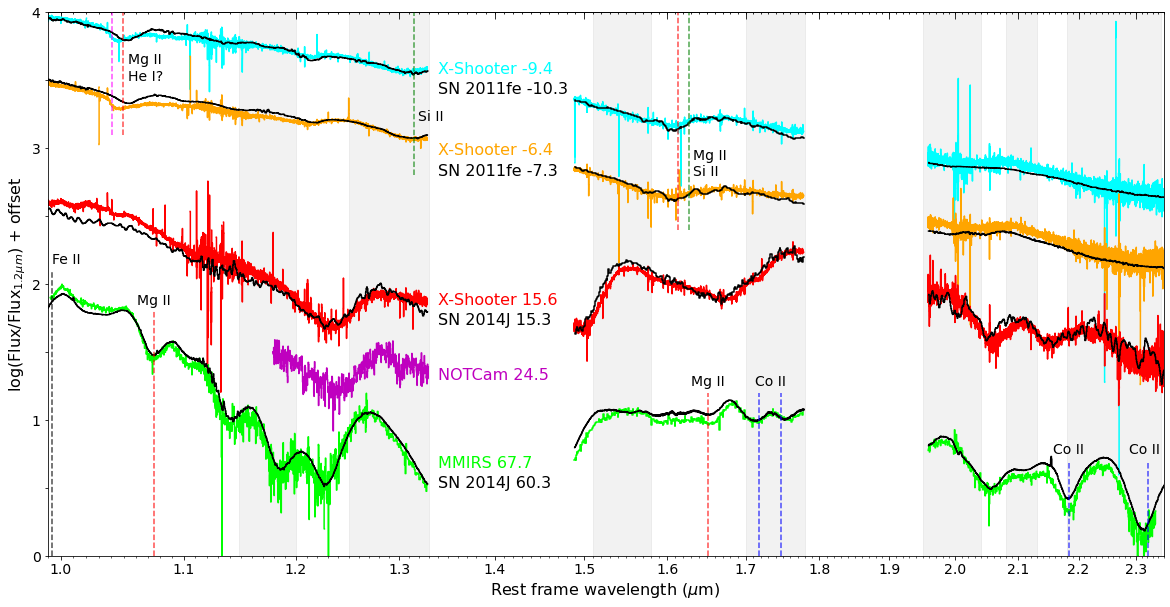}
    \caption{NIR spectra of SN~2024gy with the flux normalized at $1.2\ \mu$m for clarity. Regions with strong interference from sky lines have been cut out and the spectra are corrected for MW extinction. The vertical dashed lines mark absorption features of different species that are seen at different phases. Lines marked at the top are seen in the pre-peak spectra at $\sim$12\,000~km~s$^{-1}$, while those marked at the bottom show up later at $\sim$5\,000~km~s$^{-1}$. Spectra of SN~2011fe and SN~2014J at similar phases are overplotted in black, showing good agreement with SN~2024gy. The grey bands mark the regions that are heavily affected by line blanketing from Fe-group elements, creating a pseudo-continuum that raises the flux locally.}
    \label{NIR_specs}
\end{figure*}

\newpage
\section{Tables}
\begin{table}[h!]
    \centering
    \caption{Log of spectroscopic observations}
        \resizebox{\textwidth}{!}{
        \begin{tabular}{ccccccc}
        \hline
        \hline
        MJD &                Date \& time &  Phase (d)$^{(1)}$ &               Telescope &    Instrument & R & Wavelength range (\AA)$^{(2)}$  \\
        \hline
        60314.48 & 2024-01-05 11:36:57 & -15.2 &                     FTN &      FLOYDS-N & 550 & 3300 - 10000 \\
        60316.12 & 2024-01-07 02:51:48 & -13.8 & Three Hills Observatory & modified ALPY & 600 & 3800 - 8500 \\
        60319.48 & 2024-01-10 11:33:04 & -10.2 &                     P60 &      SEDM & 100 & 3650 - 10000 \\
        60320.35 & 2024-01-11 08:26:35 &  -9.4 &                     VLT &      X-Shooter & 5400, 8900, 5600$^{(3)}$ & 3000 - 24800 \\
        60321.25 & 2024-01-12 06:06:02 &  -8.5 &                     NOT &        ALFOSC & 360 & 3200 - 9600 \\
        60322.43 & 2024-01-13 10:20:47 &  -7.3 &                    P200 &          DBSP & 5000 & 3400 - 10500 \\
        60323.33 & 2024-01-14 07:57:36 &  -6.4 &                     VLT &      X-Shooter & 5400, 8900, 5600$^{(3)}$ & 3000 - 24800 \\
        60328.19 & 2024-01-19 04:26:33 &  -1.5 &                     NOT &        ALFOSC & 360 & 3200 - 9600 \\
        60328.30 & 2024-01-19 07:09:34 &  -1.4 &    120-mm APO refractor &      ALPY 600 & 600 & 3900 - 6900\\
        60329.09 & 2024-01-20 02:06:00 &  -0.6 & Monte Baldo Observatory &    FOSC-E5535 & 100 & 4000 - 8700\\
        60329.25 & 2024-01-20 05:58:34 &  -0.5 &                     NOT &          FIES & 67000 & 3700 - 9000\\
        60336.38 & 2024-01-27 09:09:03 &   6.7 &    120-mm APO refractor &      ALPY 600 & 600 & 3900 - 6900 \\
        60338.10 & 2024-01-29 02:23:35 &   8.4 &                     NOT &        ALFOSC & 360 & 3200 - 9600 \\
        60342.23 & 2024-02-02 05:28:19 &  12.5 &                     NOT &        ALFOSC & 360 & 3200 - 9600 \\
        60345.34 & 2024-02-05 08:09:19 &  15.6 &                     VLT &      X-Shooter & 5400, 8900, 5600$^{(3)}$ & 3000 - 24800 \\
        60352.12 & 2024-02-12 02:48:47 & 22.4 & LT & SPRAT & 350 &  4000 - 8100 \\
        60353.68 & 2024-02-13 16:14:41 &  24.0 &                    KECK &          LRIS & 500 & 3000 - 10200 \\
        60354.22 & 2024-02-14 05:21:25 &  24.5 &                     NOT &        NOTCam & 2500 & 11800 - 13300 \\
        60360.15 & 2024-02-20 03:31:39 &  30.5 &                     NOT &        ALFOSC & 360 & 3200 - 9600 \\
        60375.12 & 2024-03-06 02:50:53 &  45.4 &                     NOT &        ALFOSC & 360 & 3200 - 9600 \\
        60397.33 & 2024-03-28 07:54:00 &  67.7 &                     MMT &         MMIRS & 1800, 2600$^{(3)}$ & 9400 - 23400 \\
        60402.23 & 2024-04-02 05:30:02 &  72.5 &                     NTT &        EFOSC2 & 250, 500$^{(3)}$ & 3380 - 10320 \\
        60418.95 & 2024-04-18 22:45:59 &  89.3 &                     INT &           IDS & 1092 & 3900 - 8800 \\
        60442.13 & 2024-05-12 03:10:19 & 112.4 &                     MMT &      Binospec & 1340 & 3850 - 9150 \\
        60472.23 & 2024-06-11 05:35:36 & 142.5 &                     P60 &      SEDM & 100 & 3650 - 10000 \\
        60490.22 & 2024-06-29 05:11:35 & 160.5 &                     P60 &      SEDM & 100 & 3650 - 10000 \\
        60490.25 & 2024-06-29 06:06:34 & 160.6 &                    KECK &          LRIS & 500 & 3000 - 10200 \\
        60646.47 & 2024-12-02 11:17:19 & 316.8 &                     MMT &      Binospec & 1340 & 3850 - 9150 \\
        \hline
    \end{tabular}
    }
    \tablefoot{\\
    \tablefoottext{1}{Phase 0 is defined as the time of peak brightness $t_0 = 60329.7$, as found through \textsc{salt2} fits.\\}
    \tablefoottext{2}{Reported instrument wavelength range. The spectra may have been trimmed to remove high noise regions at the edges.\\}
    \tablefoottext{3}{Multiple spectra combined into one.}}
    \label{speclog}
\end{table}

\begin{table}[h!]
    \centering
    \caption{Host extinction estimates for the spectral matching method.}
        \resizebox{\textwidth}{!}{
        \begin{tabular}{ccccccccccc}
        \hline
        \hline
        & \multicolumn{5}{c}{SN~2011fe} & \multicolumn{5}{c}{SN~2017erp$^{(1)}$} \\
        SN 2024gy & Comparison & \multicolumn{2}{c}{$R_V=3.1$} & \multicolumn{2}{c}{$R_V=2.659$} & Comparison & \multicolumn{2}{c}{$R_V=3.1$} & \multicolumn{2}{c}{$R_V=2.659$} \\
        phase & phase & $A_V$ (mag) & \ebv$_{host}$ (mag)& $A_V$ (mag) & \ebv$_{host}$ (mag) & phase & $A_V$ (mag) & \ebv$_{host}$ (mag) & $A_V$ (mag) & \ebv$_{host}$ (mag) \\
        \hline
        $-9.4$ &       $-9.3$ & $0.75$ & $0.24$ &   $0.60$ &    $0.23$ &    $-9.6$ &  $0.60$ &   $0.19$ &     $0.50$ &      $0.19$ \\
        $-8.5$ &       $-8.1$ & $0.75$ & $0.24$ &   $0.60$ &    $0.23$ &    $-9.6$ &  $0.60$ &   $0.19$ &     $0.45$ &      $0.17$ \\
        $-7.3$ &       $-7.3$ & $0.70$ & $0.23$ &   $0.55$ &    $0.21$ &    $-7.7$ &  $0.70$ &   $0.23$ &     $0.55$ &      $0.21$ \\
        $-6.4$ &       $-6.1$ & $0.70$ & $0.23$ &   $0.60$ &    $0.23$ &    $-6.7$ &  $0.70$ &   $0.23$ &     $0.60$ &      $0.23$ \\
        $-1.5$ &       $-1.3$ & $0.70$ & $0.23$ &   $0.50$ &    $0.19$ &    $-1.7$ &  $0.45$ &   $0.15$ &     $0.40$ &      $0.15$ \\
        $-0.6$ &       $-0.3$ & $0.75$ & $0.24$ &   $0.50$ &    $0.19$ &    $ 0.4$ &  $0.60$ &   $0.19$ &     $0.50$ &      $0.19$ \\
        $8.4$ &        $8.7$ & $0.80$ & $0.26$ &   $0.65$ &    $0.24$ &        - &     - &      - &        - &         - \\
        $12.5$ &       $11.9$ & $0.75$ & $0.24$ &   $0.50$ &    $0.19$ &        - &     - &      - &        - &         - \\
        $15.6$ &       $15.9$ & $0.70$ & $0.23$ &   $0.55$ &    $0.21$ &    $16.4$ &  $0.60$ &   $0.19$ &     $0.45$ &      $0.17$ \\
        $22.4$ &       $21.9$ & $0.80$ & $0.26$ &   $0.55$ &    $0.21$ &    - &  - &   - &     - &      - \\
        $24.0^{(2)}$ &       $23.9$ & $0.70$ & $0.23$ &   $0.50$ &    $0.19$ &    $25.3$ &  $0.35$ &   $0.11$ &     $0.30$ &      $0.11$ \\
        \hline
    \end{tabular}
    }
    \tablefoot{The $A_V$ step size between comparisons is $0.05$ mag, which gives an \ebv$_{host}$ uncertainty of $0.02$ mag. The ALPY 600 and NOTCam spectra have been omitted as they do not cover the $4000\ -\ 8500$ \AA\ wavelength range, and the FIES spectrum is omitted as its high resolution prevented accurate comparisons.\\
    \tablefoottext{1}{SN~2017erp did not have comparison spectra within $\sim1$ day at all phases. The affected rows are filled with dashes.\\}
    \tablefoottext{2}{The SN is transitioning into the nebular phase. Accurately matching becomes more difficult as \ion{Fe}{} lines appear.}}
    \label{specmatch_table}
\end{table}

\begin{table}[hp!]
    \centering
    \caption{Equivalent widths and \ebv\ estimates for the fitted interstellar absorption lines.}
    \resizebox{0.69\textwidth}{!}{
    \begin{tabular}{cccccc}
        \hline
        \hline
        Spectrum$^{(1)}$ & line & rest frame wavelength (\AA) & $v$ (km s$^{-1}$)$^{(2)}$ & EW (\AA) & \ebv\ (mag) \\
        \hline
        \multirow{3}{*}{FIES} &  &  & $-24.6 \pm 1.2$ & $0.09 \pm 0.03$ &  - \\
        &  &  & $319.9 \pm 1.2$ & $0.47 \pm 0.04$ &  - \\
        &  &  & $378.3 \pm 0.9$ & $0.61 \pm 0.05$ &  - \\
        \rowcolor[gray]{.9}
        &  &  & $-10.8 \pm 1.8$ & $0.27\pm0.02$ & - \\
        \rowcolor[gray]{.9}
        \multirow{-2}{*}{X-Shooter 1} &  &  & $349.6 \pm 0.9$ & $1.03\pm0.03$ & - \\
        \multirow{2}{*}{X-Shooter 2} &  &  & $-11.1 \pm 1.5$ & $0.27\pm0.02$ & - \\
        &  &  &  $349.9 \pm 0.6$ & $1.02\pm0.02$ & - \\
        \rowcolor[gray]{.9}
        &  &  & $-12.0 \pm 2.4$ & $0.30\pm0.03$ & - \\
        \rowcolor[gray]{.9}
        \multirow{-2}{*}{X-Shooter 3} & \multirow{-9}{*}{\CaII\ K} & \multirow{-9}{*}{3933.66} &  $350.2 \pm 1.2$ & $1.03\pm0.04$ & - \\
        \hline
        \multirow{3}{*}{FIES} &  &  & $-24.6 \pm 1.2$ & $0.20 \pm 0.04$ &  - \\
        &  &  & $319.9 \pm 1.2$ & $0.31 \pm 0.04$ &  - \\
        &  &  & $378.3 \pm 0.9$ & $0.28 \pm 0.04$ &  - \\
        \rowcolor[gray]{.9}
        &  &  & $-10.8 \pm 1.8$ & $0.17\pm0.02$ & - \\
        \rowcolor[gray]{.9}
        \multirow{-2}{*}{X-Shooter 1} &  &  & $352.3 \pm 1.2$ & $0.69\pm0.03$ & - \\
        \multirow{2}{*}{X-Shooter 2} &  &  & $-11.1 \pm 1.5$ & $0.17\pm0.02$ & - \\
        &  &  & $351.7 \pm 1.2$ & $0.69\pm0.03$ & - \\
        \rowcolor[gray]{.9}
        &  &  & $-12.0 \pm 2.4$ & $0.14\pm0.02$ & - \\
        \rowcolor[gray]{.9}
        \multirow{-2}{*}{X-Shooter 3} &  \multirow{-9}{*}{\CaII\ H}  & \multirow{-9}{*}{3968.47} & $354.7 \pm 1.5$ & $0.74\pm0.04$ & - \\
        \hline
        X-Shooter 1 &  &  & $380.7 \pm 6.0$ & $0.092 \pm 0.013$ & $0.12 \pm 0.02$ \\
        \rowcolor[gray]{.9}
        X-Shooter 2 &  &  & $374.7 \pm 6.0$ & $0.092 \pm 0.011$ & $0.12 \pm 0.02$ \\
        X-Shooter 3 & \multirow{-3}{*}{DIB$_{\lambda5780}$} & \multirow{-3}{*}{5780.60} & $377.7 \pm 9.0$ & $0.085 \pm 0.015$ & $0.11 \pm 0.02$ \\
        \hline
        X-Shooter 1 &  &  & $353.8 \pm 15.0$ & $0.022 \pm 0.011$ & $0.08 \pm 0.07$ \\
        \rowcolor[gray]{.9}
        X-Shooter 2 &  &  & $374.7 \pm 3.0$ & $0.036 \pm 0.007$ & $0.13 \pm 0.07$ \\
        X-Shooter 3 & \multirow{-3}{*}{DIB$_{\lambda5797}$} & \multirow{-3}{*}{5797.10} & $365.7 \pm 15.0$ & $0.036 \pm 0.012$ & $0.13 \pm 0.08$ \\
        \hline
        \multirow{3}{*}{FIES} &  &  & $-24.6 \pm 1.2$ & $0.27\pm0.07$ & $0.05\pm0.02$ \\
        &  &  & $319.9 \pm 1.2$ & $0.66\pm0.08$ & $0.34\pm0.18$ \\
        &  &  & $378.3 \pm 0.9$ & $0.84\pm0.12$ & $0.79\pm0.54$ \\
        \rowcolor[gray]{.9}
        &  &  & $-23.7 \pm 1.5$ & $0.25\pm0.03$ & $0.04\pm0.02$ \\
        \rowcolor[gray]{.9}
        &  &  & $321.1 \pm 0.9$ & $0.63\pm0.04$ & $0.28\pm0.11$ \\
        \rowcolor[gray]{.9}
        \multirow{-3}{*}{X-Shooter 1} &  &  & $377.1 \pm 0.6$ & $0.82\pm0.04$ & $0.74\pm0.29$ \\
        \multirow{3}{*}{X-Shooter 2} &  &  & $-22.5 \pm 1.5$ & $0.26\pm0.03$ & $0.05\pm0.02$ \\
        &  &  & $321.7 \pm 0.9$ & $0.62\pm0.04$ & $0.27\pm0.11$ \\
        &  &  & $377.4 \pm 0.9$ & $0.83\pm0.04$ & $0.77\pm0.30$ \\
        \rowcolor[gray]{.9}
        &  &  & $-22.8 \pm 0.9$ & $0.23\pm0.02$ & $0.04\pm0.01$ \\
        \rowcolor[gray]{.9}
        &  &  & $321.7 \pm 0.6$ & $0.62\pm0.02$ & $0.27\pm0.10$ \\
        \rowcolor[gray]{.9}
        \multirow{-3}{*}{X-Shooter 3} & \multirow{-12}{*}{\NaI\ D$_2$} & \multirow{-12}{*}{5889.95} & $378.0 \pm 0.3$ & $0.83\pm0.02$ & $0.77\pm0.28$ \\
        \hline
        \multirow{3}{*}{FIES} &  &  & $-24.6 \pm 1.2$ & $0.17\pm0.07$ & $0.05\pm0.02$ \\
        &  &  & $319.9 \pm 1.2$ & $0.55\pm0.08$ & $0.40\pm0.24$ \\
        &  &  & $378.3 \pm 0.9$ & $0.72\pm0.10$ & $1.05\pm0.75$ \\
        \rowcolor[gray]{.9}
        &  &  & $-23.7 \pm 1.5$ & $0.16\pm0.03$ & $0.04\pm0.02$ \\
        \rowcolor[gray]{.9}
        &  &  & $321.1 \pm 0.9$ & $0.49\pm0.03$ & $0.27\pm0.12$ \\
        \rowcolor[gray]{.9}
        \multirow{-3}{*}{X-Shooter 1} &  &  & $377.1 \pm 0.6$ & $0.73\pm0.04$ & $1.12\pm0.49$ \\
        \multirow{3}{*}{X-Shooter 2} &  &  & $-22.5 \pm 1.5$ & $0.17\pm0.03$ & $0.05\pm0.02$ \\
        &  &  & $321.7 \pm 0.9$ & $0.48\pm0.03$ & $0.27\pm0.12$ \\
        &  &  & $377.4 \pm 0.9$ & $0.73\pm0.03$ & $1.10\pm0.48$ \\
        \rowcolor[gray]{.9}
        &  &  & $-22.8 \pm 0.9$ & $0.15\pm0.02$ & $0.04\pm0.02$ \\
        \rowcolor[gray]{.9}
        &  &  & $321.7 \pm 0.6$ & $0.50\pm0.02$ & $0.29\pm0.12$ \\
        \rowcolor[gray]{.9}
        \multirow{-3}{*}{X-Shooter 3} & \multirow{-12}{*}{\NaI\ D$_1$} & \multirow{-12}{*}{5895.92} & $378.0 \pm 0.3$ & $0.74\pm0.02$ & $1.16\pm0.48$ \\
        \hline
        X-Shooter 1 &  &  & $398.7 \pm 33.0$ & $0.11 \pm 0.03$ &  - \\
        \rowcolor[gray]{.9}
        X-Shooter 2 &  &  & $596.6 \pm 24.0$ & $0.29 \pm 0.03$ &  - \\
        X-Shooter 3 & \multirow{-3}{*}{DIB$_{\lambda6275}$} & \multirow{-3}{*}{6275.60} & $425.7 \pm 63.0$ & $0.14 \pm 0.05$ &  - \\
        \hline
        X-Shooter 1 &  &  & $392.7 \pm 9.0$ & $0.16 \pm 0.02$ &  - \\
        \rowcolor[gray]{.9}
        X-Shooter 2 &  &  & $392.7 \pm 6.0$ & $0.08 \pm 0.01$ &  - \\
        X-Shooter 3 & \multirow{-3}{*}{DIB$_{\lambda6284}$} & \multirow{-3}{*}{6284.10} & $380.7 \pm 6.0$ & $0.13 \pm 0.02$ &  - \\
        \hline
        X-Shooter 1 &  &  & $398.7 \pm 9.0$ & $0.032 \pm 0.008$ & $0.17 \pm 0.17$ \\
        \rowcolor[gray]{.9}
        X-Shooter 2 &  &  & $389.7 \pm 6.0$ & $0.047 \pm 0.007$ & $0.26 \pm 0.24$ \\
        X-Shooter 3 & \multirow{-3}{*}{DIB$_{\lambda6613}$} & \multirow{-3}{*}{6613.70} & $368.7 \pm 3.0$ & $0.043 \pm 0.007$ & $0.24 \pm 0.22$ \\
        \hline
        \multirow{2}{*}{X-Shooter 1} &  &  & $312.1 \pm 4.5$ & $0.017 \pm 0.008$ &   - \\
        &  &  & $372.3 \pm 0.6$ & $0.127 \pm 0.007$ &   - \\
        \rowcolor[gray]{.9}
         &  &  & $307.6 \pm 1.8$ & $0.035 \pm 0.011$ &   - \\
        \rowcolor[gray]{.9}
        \multirow{-2}{*}{X-Shooter 2}&  &  & $372.0 \pm 1.2$ & $0.135 \pm 0.015$ &   - \\
        \multirow{2}{*}{X-Shooter 3} &  &  & $310.3 \pm 7.2$ & $0.016 \pm 0.010$ &   - \\
        & \multirow{-6}{*}{\KI$_1$} & \multirow{-6}{*}{7664.90} & $372.3 \pm 0.6$ & $0.115 \pm 0.008$ &   - \\
        \hline
        X-Shooter 1 &  &  & $372.3 \pm 0.6$ & $0.071 \pm 0.005$ & $0.26 \pm 0.15$ \\
        \rowcolor[gray]{.9}
        X-Shooter 2 &  &  & $372.0 \pm 1.2$ & $0.069 \pm 0.011$ & $0.26 \pm 0.15$ \\
        X-Shooter 3 & \multirow{-3}{*}{\KI$_2$} & \multirow{-3}{*}{7698.97} & $372.3 \pm 0.6$ & $0.068 \pm 0.006$ & $0.25 \pm 0.15$ \\
        \hline
    \end{tabular}
    }
    \tablefoot{The first column shows the spectrum that was used to fit the line, the second and third columns give the line name and rest frame wavelength. Column 4 gives the velocity of the fitted line in the observer frame, and column 5 its equivalent width. The last column shows the resulting \ebv\ estimate. The relations used for these estimates come from  \citet{KI_EBV_relation} (\KI$_2$), \citet{NaID_EBV_relation} (\NaI\ D), and \citet{DIB_EBV_relation} (DIBs).\\
    \tablefoottext{1}{X-Shooter 1, 2, and 3 are the spectra taken on 11 Jan, 14 Jan, and 5 Feb, respectively.\\}
    \tablefoottext{2}{Measured recession velocity in the observer frame. For comparison, $v_{2024gy} = 354.7$ km s$^{-1}$ in this frame.}
    }
    \label{EWs}
\end{table}

\end{appendix}
\end{document}